\documentclass[letterpaper,twocolumn,10pt]{article}
\usepackage{usenix}

\usepackage{tikz}
\usepackage{amsmath}

\usepackage{filecontents}

\usepackage{cite}
\usepackage{amsmath,amssymb,amsfonts}
\usepackage{algorithmic}
\usepackage{graphicx}
\usepackage{textcomp}
\usepackage{xcolor}
\usepackage{booktabs}
\usepackage{tabularx}
\usepackage{multirow}
\usepackage{subcaption}
\usepackage{url}
\usepackage{adjustbox}
\usepackage{enumitem}
\usepackage{makecell}  % Add this to your preamble if not already present

\begin{document}

%don't want date printed
\date{}

\title{\Large \bf Evasion of IoT Malware Detection via Dummy Code Injection }

\author{
Sahar Zargarzadeh \\
University of Texas at Arlington \\
\texttt{sxz3379@mavs.uta.edu}
\and
Mohammad A. Islam \\
University of Texas at Arlington \\
\texttt{mislam@uta.edu}
}

\maketitle

\begin{abstract}
The Internet of Things (IoT) has revolutionized connectivity by linking billions of devices worldwide. However, this rapid expansion has also introduced severe security vulnerabilities, making IoT devices attractive targets for malware such as the Mirai botnet. Power side-channel analysis has recently emerged as a promising technique for detecting malware activity based on device power consumption patterns. However, the resilience of such detection systems under adversarial manipulation remains underexplored.

This work presents a novel adversarial strategy against power side-channel-based malware detection. By injecting structured dummy code into the scanning phase of the Mirai botnet, we dynamically perturb power signatures to evade AI/ML-based anomaly detection without disrupting core functionality. Our approach systematically analyzes the trade-offs between stealthiness, execution overhead, and evasion effectiveness across multiple state-of-the-art models for side-channel analysis, using a custom dataset collected from smartphones of diverse manufacturers. Experimental results show that our adversarial modifications achieve an average attack success rate of 75.2\%, revealing practical vulnerabilities in power-based intrusion detection frameworks.
\end{abstract}

\section{Introduction}
\begin{figure}[!t]
    \centering
    \begin{minipage}{0.49\linewidth} 
        \centering
        \includegraphics[width=\linewidth]{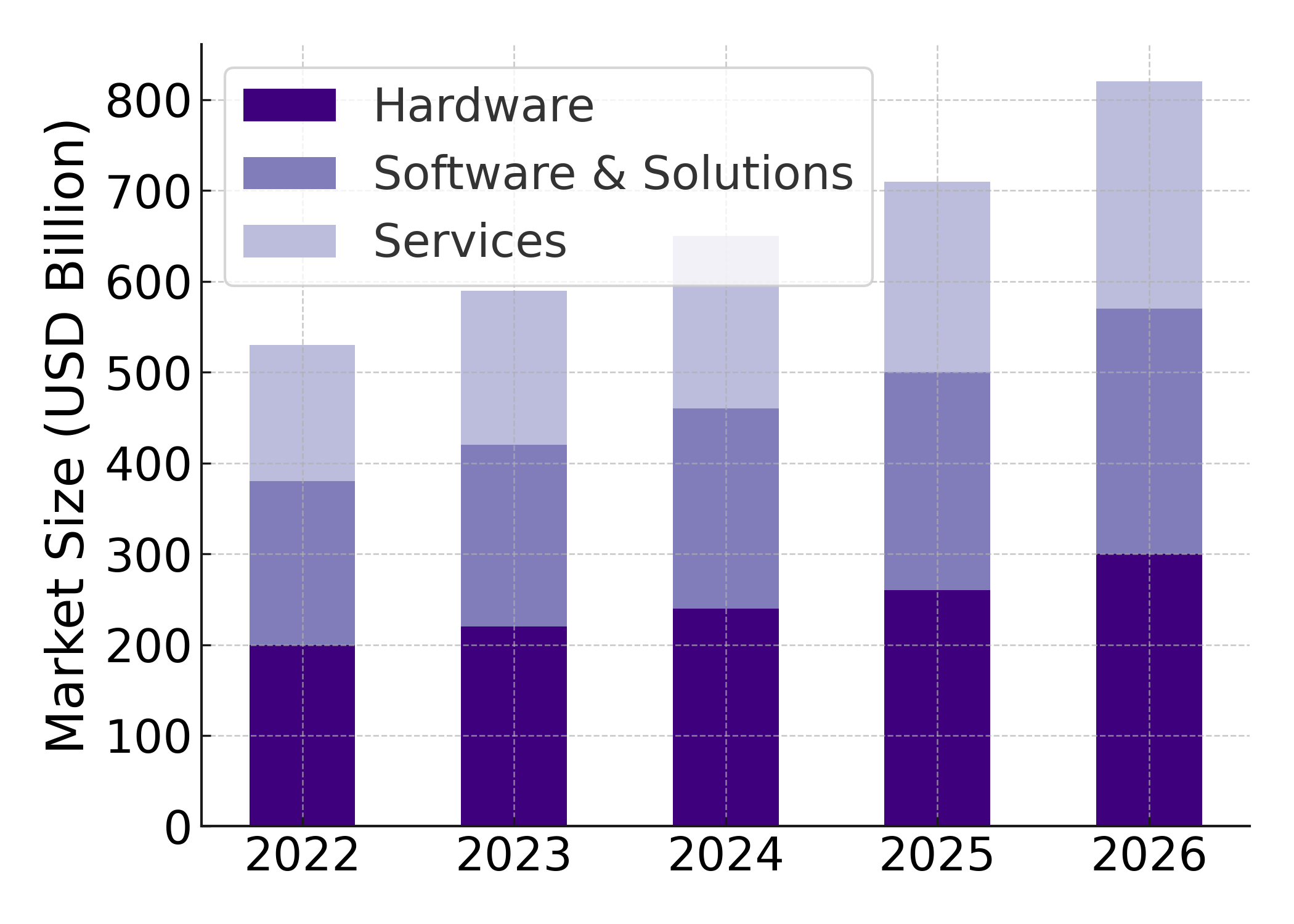}
    \end{minipage}
    \hfill
    \begin{minipage}{0.49\linewidth}
        \centering
        \includegraphics[width=\linewidth]{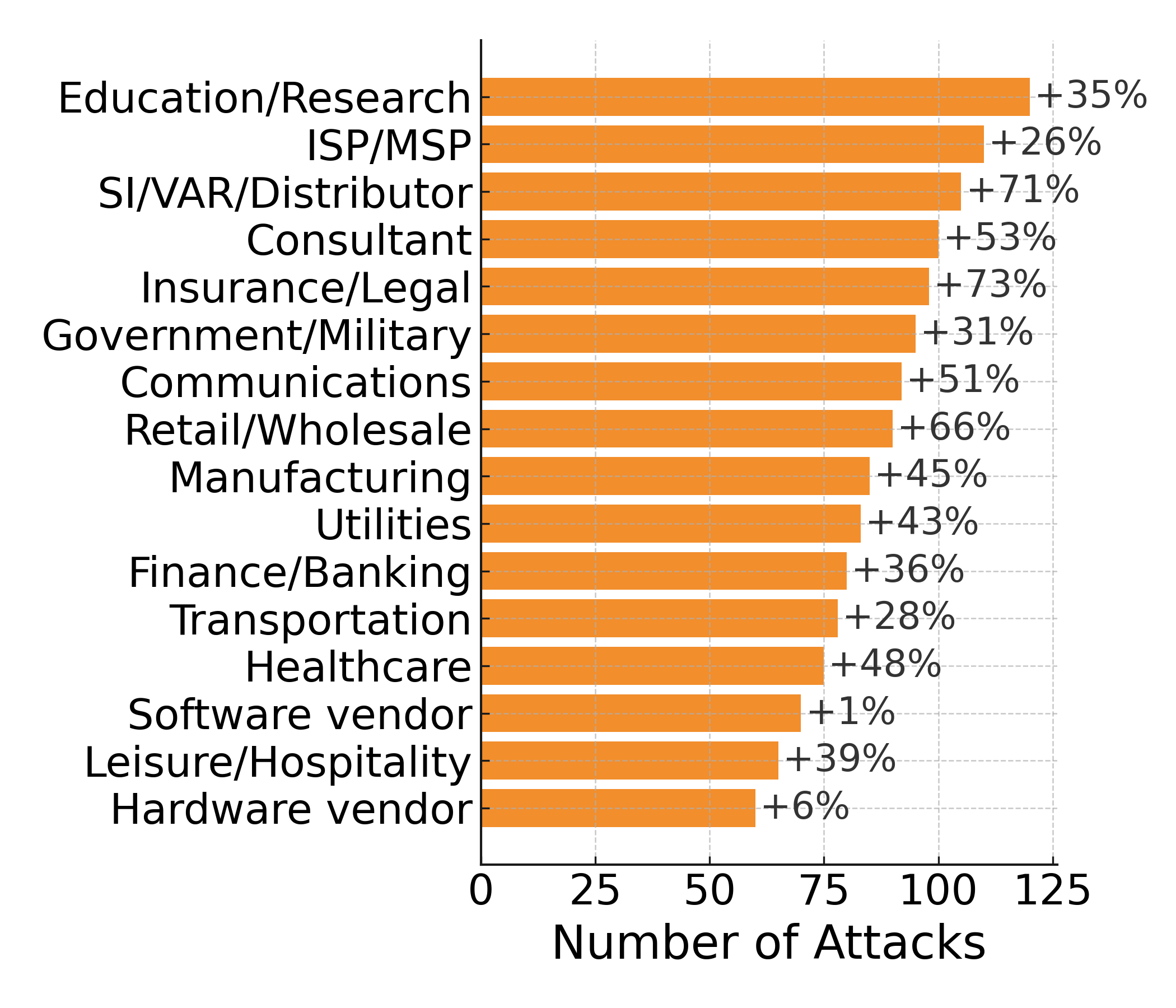}
    \end{minipage}
   \caption{(a) Global enterprise IoT market growth from 2022 to 2026 (projected)~\cite{IoTMarket2023}. 
(b) Average weekly IoT cyberattacks by sector in Jan–Feb 2023~\cite{IoTCyberAttacks2023}. }

    \label{fig:IoT_Threat_Market_Trend}
\end{figure}

The Internet of Things (IoT) has seamlessly integrated billions of devices into personal, industrial, and critical infrastructure domains. Despite the transformative potential, IoT devices remain highly vulnerable to cyber threats due to their constrained resources, lack of standardized security mechanisms, and outdated firmware. Recent trends show a dramatic increase in the number of connected devices and corresponding growth in cyberattack frequency, as illustrated in Figure~\ref{fig:IoT_Threat_Market_Trend}. 
The Mirai botnet attack of 2016 demonstrated how weakly secured IoT devices could be weaponized to launch large-scale Distributed Denial of Service (DDoS) attacks, exposing systemic flaws in global IoT security~\cite{antonakakis2017understanding}. Since then, variants of Mirai have continued to evolve, leveraging the proliferation of insecure devices to sustain botnet-driven threats across diverse sectors~\cite{affinito2023evolution}.

\textbf{Motivation.} 
As IoT adoption accelerates across critical sectors, the attack surface for malware has expanded significantly. Adaptive threats such as evolving variants of the Mirai botnet are increasingly capable of evading conventional detection systems that rely on static signatures or network traffic analysis. These traditional techniques are often ineffective against obfuscated malware that manipulates communication behavior to bypass intrusion detection systems.

Side-channel analysis, developed to extract cryptographic secrets via power or timing signals~\cite{lumbiarres2016hardware}, has recently been repurposed as a defensive mechanism in IoT security. By monitoring device-level power consumption, researchers have shown that malware activity induces distinctive runtime signatures that can be used for anomaly detection~\cite{jung2020iot, cathis2024sok, ding2020deeppower, jung2022deepauditor}. Unlike host- or network-based approaches, power side-channel analysis is non-intrusive, operates out-of-band, and is inherently more difficult for malware to tamper with.

This shift from attack vector to defense mechanism has led to the development of detection systems such as DeepPower~\cite{ding2020deeppower} and DeepAuditor~\cite{jung2022deepauditor}, which leverage machine learning to classify behavioral patterns in power traces. Systematization-of-Knowledge (SoK) efforts~\cite{cathis2024sok} highlight both the promise and the limitations of this approach, particularly concerns around generalization and robustness when facing adaptive threats. Recent studies, including our own, indicate that despite their vantage point, power-based detectors remain susceptible to evasion by sophisticated attackers.

Power-based anomaly detection relies on the assumption that different computational states, including those associated with malware, produce distinguishable power signatures. While ML-based side-channel systems have demonstrated strong performance under clean conditions, their resilience to targeted runtime perturbations remains underexplored. Our work addresses this gap by demonstrating how adversarial control over execution phases can strategically disrupt power signatures to evade detection.

Inspired by traditional countermeasures in cryptographic contexts such as masking, hiding, and noise injection, which reduce leakage by perturbing power characteristics~\cite{power_amplitude_increase}, we invert this defensive principle. Rather than using perturbations to conceal secrets, we introduce structured modifications to deceive AI-based malware detectors. Specifically, we inject dummy instructions into Mirai’s scanning phase to desynchronize power traces in a controlled, targeted manner~\cite{gao2024deeptheft}.

This form of adversarial desynchronization exploits a core weakness in current detection models: their reliance on fixed-phase patterns learned during training~\cite{qiu2024power}. By subtly manipulating power traces to resemble benign noise, without altering core malware functionality, we achieve stealthy, model-agnostic evasion. Our design leverages both the attacker’s control over execution timing and the sensitivity of side-channel classifiers to temporal variation, revealing a novel and practical evasion vector.

\textbf{Our contributions.} This work makes the following contributions:

\begin{itemize}
\item \textbf{Adversarial Evasion Strategy:} We propose a novel runtime perturbation technique that renders Mirai botnet activity undetectable to power-based anomaly detectors. Our approach injects structured dummy code during the scanning phase to manipulate power signatures in a targeted and stealthy manner, guided by explainable AI (SHAP) to disrupt critical features.

\item \textbf{Power Side-Channel Dataset:} We collect a real-world dataset comprising power consumption traces from multiple smartphone platforms under four operational states: idle, IoT service, unmodified Mirai, and modified Mirai. The dataset captures both benign and adversarial behaviors in realistic settings.

\item \textbf{Cross-Architecture Evaluation:} We evaluate our evasion method across six deep learning-based detection architectures, including sequential models, Long Short-Term Memory (LSTM) and Bidirectional LSTM (BiLSTM); convolutional models, Temporal Convolutional Network (TCN) and BiLSTM combined with one-dimensional Convolutional Neural Network (BiLSTM+CNN); and reconstruction-based models, CNN with self-attention (CNN+Attention) and LSTM Autoencoder with Multi-Layer Perceptron (AE+MLP). Results demonstrate successful evasion with attack success rates exceeding 75\% across most models.

\item \textbf{Trade-off Characterization:} We analyze the relationship between evasion success, stealthiness, and execution overhead using statistical tools including Pearson correlation, ANOVA, and Granger causality. These findings provide insights into the AI/ML design space and inform future defense strategies.
\end{itemize}

Our results show that power-based side-channel detection systems, particularly those using sequential models like LSTM, can be reliably bypassed, with attack success rates (ASR) exceeding 75\% in many cases. This exposes a key vulnerability in detectors that assume static power signatures~\cite{cathis2024sok}. Although our evaluation focuses on mobile platforms and the Mirai malware family, these choices reflect real-world relevance: smartphones are high-value mobile IoT targets, and Mirai remains a widely used and evolving botnet~\cite{antonakakis2017understanding, affinito2023evolution}. As a representative case, Mirai’s scanning behavior allows our findings to generalize to other propagation-based malware.

\begin{figure*}[t!]
\centerline{\includegraphics[width=12cm]{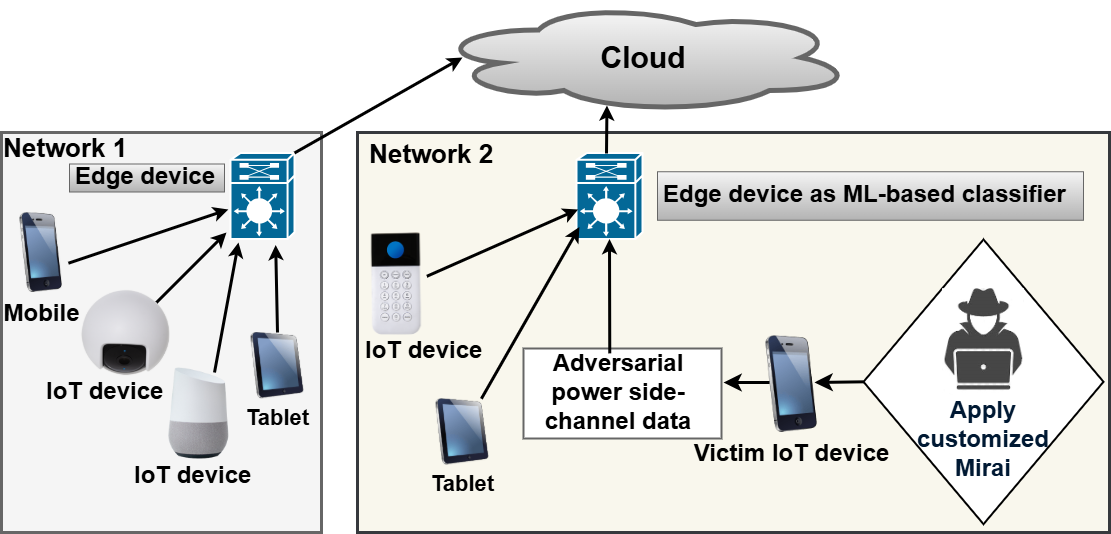}}
\caption{System architecture overview showing IoT devices, edge nodes, and cloud-based anomaly detection pipeline.}
\label{fig:FirstNetwork}
\end{figure*}

\section{Background}
\label{sec:related_work}

\subsection{Adversarial Machine Learning}
Adversarial Machine Learning (AML) studies how deliberately crafted perturbations can manipulate the behavior of machine learning models~\cite{Biggio2013Evasion, szegedy2013intriguing, carlini2017adversarial}. Early research primarily targeted domains such as image recognition and natural language processing, where adversarial inputs often manifest as imperceptible pixel changes or subtle textual modifications. More recently, AML techniques have been extended to physical-world settings, including sensor spoofing~\cite{cao2019adversarial, kurakin2018adversarial} and adversarial perturbations on time-series sensor data, such as LiDAR or acoustic signals~\cite{sun2020towards}, revealing that models trained on hardware-derived measurements are equally susceptible.

Side-channels such as power consumption, electromagnetic (EM) emissions, and acoustic signals present unique challenges for adversarial manipulation due to their temporal structure, measurement noise, and hardware-dependent variability~\cite{athalye2018obfuscated, horvath2024sok}. While the adversarial threat model is well-studied in computer vision, its adaptation to side-channel contexts remains underexplored. This gap is significant because side-channel–based detection systems are increasingly deployed in resource-constrained IoT devices, where traditional security measures may be infeasible.

Historically, side-channel analysis has been extensively used in cryptography, where techniques such as Differential Power Analysis (DPA) can extract cryptographic keys from power traces~\cite{power_amplitude_increase}. These same principles correlate physical measurements with internal computations and underpin recent work applying deep learning to detect malicious behavior via power signals.

\subsection{Power Side-Channel}
IoT devices exhibit distinct power consumption patterns depending on their operational state. Leveraging this property, recent research has explored side-channel–based malware detection, where power traces serve as behavioral fingerprints. Unlike network-based detection, this approach remains effective even when malicious code is obfuscated, encrypted, or executed in-memory without disk artifacts~\cite{ding2020deeppower, jung2022deepauditor}.

Systems such as DeepPower~\cite{ding2020deeppower} and DeepAuditor~\cite{jung2022deepauditor} have demonstrated that deep learning models can distinguish between benign and malicious workloads on IoT devices with high accuracy. These models exploit fine-grained temporal variations such as periodicity, amplitude shifts, and transient spikes linked to specific malware phases such as scanning or attack execution. However, most prior work assumes a \emph{passive} adversary and overlooks adaptive strategies that can intentionally perturb these power signatures. Emerging studies on adversarial examples for time-series data~\cite{fawaz2019adversarial, harford2020adversarial} suggest that such perturbations could be designed to evade even state-of-the-art detectors.

Our work addresses this overlooked gap by designing targeted runtime perturbations that disrupt the structured features learned by side-channel models. Specifically, we inject carefully structured dummy code into malware execution to alter its side-channel footprint without impairing functionality, directly challenging the assumptions of existing detection frameworks.

\subsection{Mirai Botnet}
The Mirai botnet~\cite{antonakakis2017understanding} remains one of the most influential examples of IoT-targeting malware, notable for its role in record-breaking DDoS attacks. Mirai propagates by scanning the IPv4 address space for vulnerable devices, exploiting default credentials or hard-coded passwords to gain access. Once compromised, devices join a command-and-control (C2) network to await attack instructions. The scanning phase, characterized by rapid connection attempts and lightweight credential brute-forcing, produces distinct and repeatable power consumption patterns.

Since its public release, Mirai has inspired a proliferation of variants~\cite{affinito2023evolution, kolias2017ddos}, each introducing changes to propagation logic, target selection, and attack payloads. This adaptability poses challenges for defenders and underscores the importance of understanding dynamic malware behavior. Because scanning-based propagation is common in other IoT malware families~\cite{li2022thingnet, pa2015iotpot}, our perturbation-based evasion approach generalizes beyond Mirai.

By focusing on Mirai’s scanning phase as the perturbation target, we demonstrate that even minor runtime modifications, such as strategically placed dummy loops or conditionals, can significantly alter the side-channel signature, enabling evasion of detection systems without diminishing the malware's operational effectiveness.

% This paper is organized as follows. Section~\ref{sec:attack_model} introduces the threat model, detailing attacker capabilities, constraints, and the dummy code injection strategy. Section~\ref{sec:performance_evaluation} describes the experimental methodology and power measurement setup, and presents the evaluation results, highlighting the impact of perturbations on multiple state-of-the-art detection models. Section~\ref{sec:discussion} examines the implications for defenders and outlines potential countermeasures, while Section~\ref{sec:conclusion} offers concluding remarks.

\section{Attack Framework}
\label{sec:attack_model}
\begin{figure*}[!t]
\centering
\includegraphics[width=0.8\textwidth, keepaspectratio]{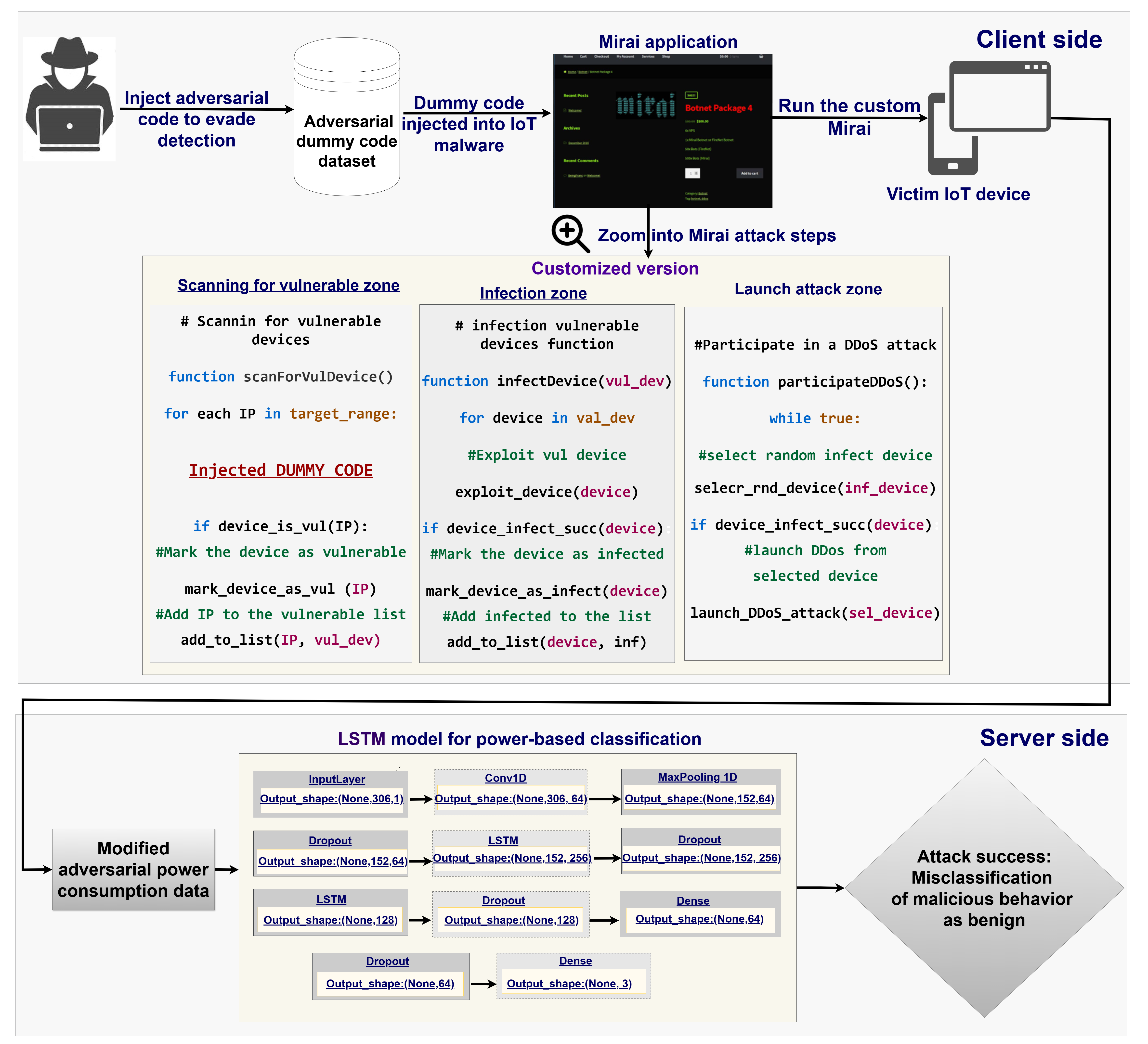}
\caption{Attack framework illustrating dummy code injection at the client side and anomaly detection at the server side.}
\label{fig:AttackStructure}
\end{figure*}
This section details our methodology for evading AI/ML-based intrusion detection systems that leverage power side-channel monitoring. Focusing on the scanning phase of the Mirai botnet, we demonstrate how targeted runtime perturbations can manipulate power consumption patterns to induce misclassification.
% Our findings expose critical vulnerabilities in power-based anomaly detection systems, revealing practical risks even for state-of-the-art models.

\subsection{Threat Model}
\label{sec:threat_model}

\textbf{System overview.}  
Our detection framework is structured as a three-tier architecture consisting of IoT devices, edge nodes, and a cloud backend, as illustrated in Figure~\ref{fig:FirstNetwork}. IoT devices continuously monitor and transmit power consumption data via Wi-Fi to nearby edge nodes. These edge nodes host lightweight machine learning classifiers that analyze incoming power traces in real time to detect malware activity, such as Mirai infections, based on side-channel signatures. The cloud backend is responsible for managing the lifecycle of detection models, including periodic retraining and distribution of updated model parameters to edge nodes. While all anomaly detection is performed locally at the edge to minimize latency and data exposure, the cloud provides centralized orchestration, model updates, and long-term storage for aggregated data and forensic analysis. This design balances the need for rapid, device-level detection with the adaptability and scalability provided by cloud-based model management, ensuring the system can respond efficiently to evolving malware threats and diverse IoT deployments~\cite{lightbody2024dragon_pi}

\textbf{Attack objectives.}  
The primary objective of the adversary is to evade power-based anomaly detection systems during the execution of Mirai malware. Specifically, the attacker aims to alter the power consumption signature of Mirai’s scanning phase so that it closely resembles benign device behavior. This misrepresentation is intended to deceive machine learning-based detectors into classifying malicious activity as normal. A successful attack results in the undetected propagation of the malware while preserving its scanning functionality and operational goals.

\textbf{Adversary capabilities and knowledge.}  
We consider a gray-box adversary who has access to the Mirai source code and is capable of modifying it before deployment. The adversary can implant dummy operations to perturb the malware’s power signature without affecting its core functionality. We assume the attacker is aware that detection is based on time-series power consumption data but does not have access to the specific model internals, such as parameters, training data, or exact thresholds. This reflects a realistic setting, given the widespread availability of Mirai’s source code and public documentation of power-based anomaly detection techniques in recent literature~\cite{jung2020iot, ding2020deeppower, cathis2024sok}.

\textbf{Adversary limitations.}  
While the attacker modifies Mirai’s scanning logic to evade detection, they deliberately preserve its core payload behavior, such as post-infection DDoS execution, to maintain the malware’s intended functionality. The attacker cannot tamper with the power monitoring infrastructure employed by the detector, including both hardware-based power tracing components and their associated data logging mechanisms. Additionally, the attacker lacks access to the detector’s internal model parameters, training data, or backend configuration. These constraints define a realistic gray-box threat model centered on software-level evasion without privileged access to device internals or the detection pipeline.

\textbf{Positioning relative to prior work.}  
This threat model differs from prior side-channel attack studies (e.g.,~\cite{chen2021voltpillager}) that assume physical access or firmware modification capabilities. Our adversary operates remotely at the application layer, focusing on software-level perturbations feasible in real-world IoT malware deployment scenarios. Unlike white-box threat models~\cite{zhang2023adversarial} assuming full access to detection models, we consider a gray-box adversary constrained to indirect profiling and pre-infection malware modification.

\textbf{Mirai signature targeting.}  The classifier distinguishes among three device states: Idle, IoT Service, and Mirai Botnet. Among these, the Mirai scanning phase, responsible for identifying vulnerable targets—is particularly computation, heavy and produces structured, high-frequency power consumption patterns that are distinct from benign behavior~\cite{golder2019practical}. Anomaly detection models leverage these patterns to differentiate Mirai activity from normal device operation.

To circumvent this detection capability, our attack targets the scanning phase of Mirai. This phase is ideal because it is both detectable and essential to malware propagation. Modifying it enables power signature perturbation without disrupting Mirai’s ability to function. To achieve this, we introduce a software-level evasion strategy that perturbs scanning-phase behavior through the injection of benign instructions. These modifications alter the power signature in a controlled manner, masking key features used by AI/ML classifiers~\cite{jung2022deepauditor}.

\subsection{Proposed Attack Strategy}
\label{sec:attack_strategy}

To evade power-based anomaly detection, we propose a runtime perturbation strategy that introduces dummy code into the Mirai malware’s scanning logic. Dummy code consists of benign computational instructions that produce structured variations in power consumption without interfering with the malware’s propagation. This approach is attractive because it enables fine-grained, software-level manipulation of power signatures while preserving Mirai’s core functionality. By targeting the most discriminative regions of the power trace, this method degrades classifier accuracy without triggering functional anomalies detectable by traditional defenses.

On the client side (i.e., the compromised IoT device), we embed different variants of dummy code into the scanning routine to generate subtle yet targeted perturbations in the power trace. These modifications are designed to degrade the performance of anomaly detection models by disrupting temporal regularities in power consumption that distinguish benign from malicious behavior.
On the server side (e.g., edge and cloud infrastructure), various AI/ML classifiers process the incoming power measurements to detect anomalies. While our evaluation includes six different architectures, Figure~\ref{fig:AttackStructure} illustrates a representative example using a hybrid BiLSTM+CNN model. This enables assessment of our attack’s generalizability across diverse detection approaches.
\begin{figure}[ht]
    \centering
    \begin{subfigure}[b]{0.44\textwidth}
        \centering
        \includegraphics[width=\textwidth]{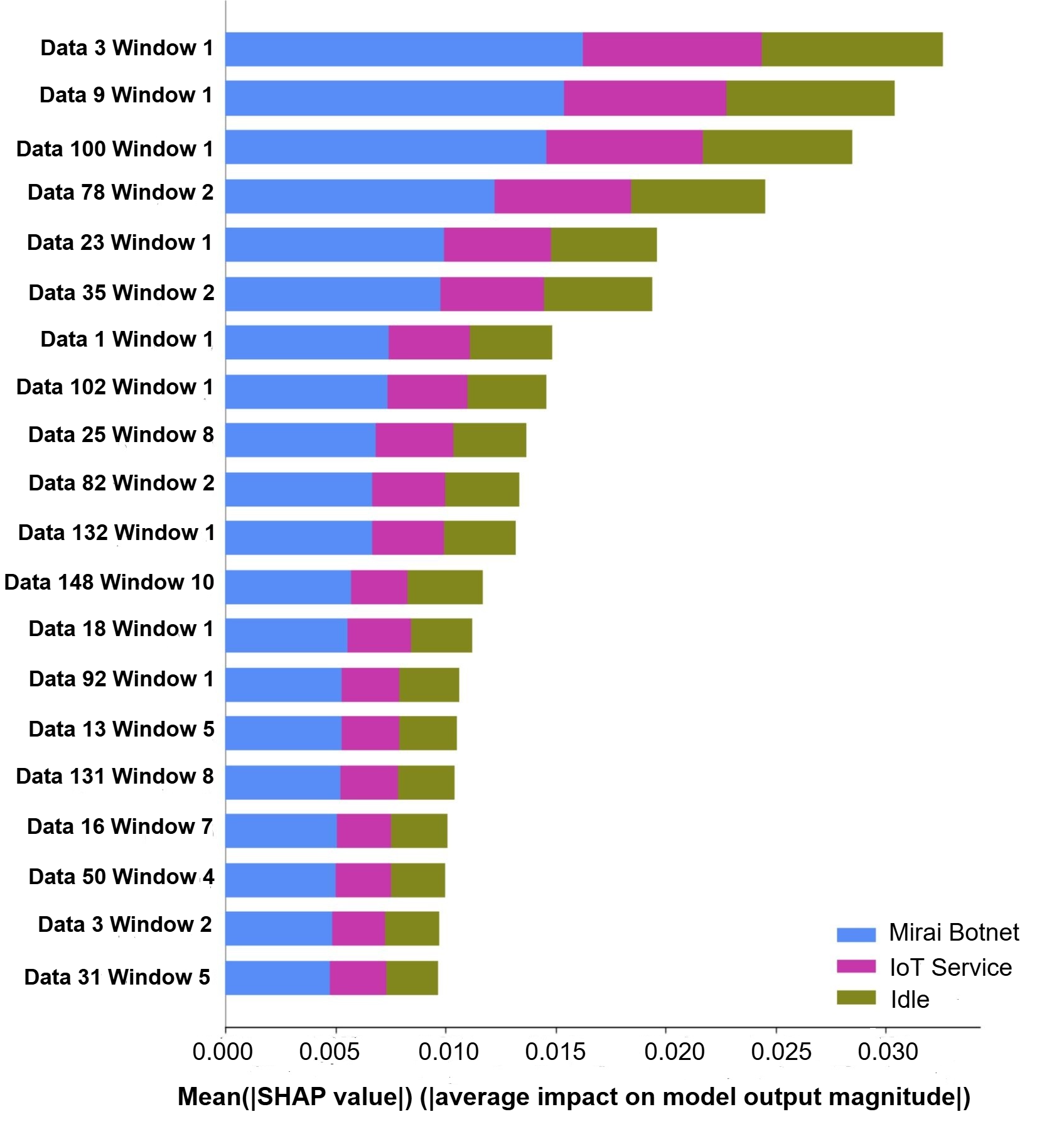}
        \caption{SHAP impact across classes (Idle, IoT Service, Mirai).}
        \label{fig:SHAP_Importance}
    \end{subfigure}
    \hfill
    \begin{subfigure}[b]{0.48\textwidth}
        \centering
        \includegraphics[width=\textwidth]{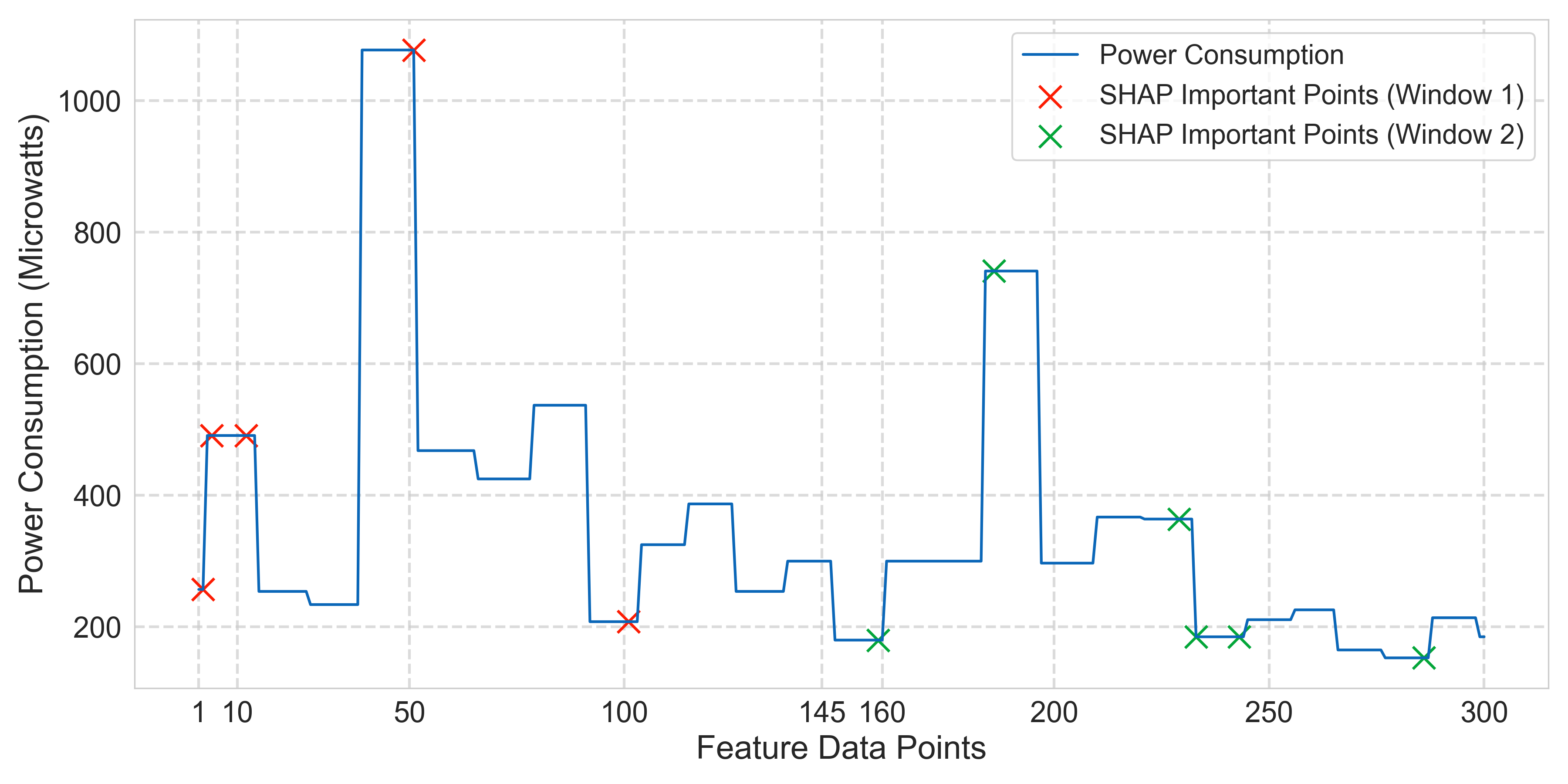}
        \caption{Power traces overlaid with SHAP-important features.}
        \label{fig:Power_Consumption_SHAP}
    \end{subfigure}
    \caption{(A) SHAP feature importance values across classes: Idle, IoT Service, and Mirai. (B) Power signal alignment with SHAP-critical data points.}
    \label{fig:SHAP_and_Power}
\end{figure}

\subsection{Feature Importance Analysis}
\label{sec:feature_importance}

To gain deeper insight into the susceptibility of AI/ML-based power side-channel detectors, we first determine which features exert the most significant influence on classification outcomes. We apply SHapley Additive exPlanations (SHAP) to quantify per-feature contributions for each class label, following the methodology in~\cite{baniecki2024adversarial}. This interpretability framework enables a direct mapping between high-impact features and distinct temporal segments in the power trace.

Figure~\ref{fig:SHAP_and_Power}(a) presents the mean absolute SHAP values for the three classes—Idle, IoT Service, and Mirai Botnet—aggregated across multiple data windows. High SHAP scores indicate features that, when altered, most significantly shift the model’s decision boundaries. The results reveal that the Mirai class is strongly associated with specific temporal windows corresponding to scanning-phase activity, which produces distinctive high-amplitude fluctuations in power consumption. To validate this relationship, Figure~\ref{fig:SHAP_and_Power}(b) overlays the SHAP-identified important points on the original power consumption trace. The visual alignment confirms that SHAP-critical features coincide with sharp power peaks and transitions, particularly those linked to Mirai’s aggressive probing behavior. These peaks represent discriminative cues that detection models leverage to separate malicious from benign patterns.

This analysis directly informs the adversarial perturbation strategy: by locating SHAP-high regions in the power signal, we can strategically target and modify the most influential temporal segments during Mirai’s execution. In our framework, dummy code injection is timed to coincide with these high-impact points, thereby flattening or distorting peak structures without disrupting core malware functionality. Such targeted modifications aim to degrade the discriminative utility of these features, effectively reducing classifier confidence and increasing the probability of evasion across diverse detection architectures.

\subsection{Generation of Dummy Code Dataset}
\label{sec:dummy_code_generation}

This section details the generation, injection, and analysis of dummy code variants designed to perturb Mirai’s power consumption behavior while preserving malware functionality. The effectiveness of each dummy code variant is evaluated through statistical power analysis and detection performance metrics.

\begin{table*}[ht]
\caption{Comparison of dummy code variants based on statistical indicators of adversarial impact.}
\label{table:DummyCodeAnalysis}
\centering
\begin{adjustbox}{width=\textwidth}
\begin{tabular}{lccccc}
\toprule
\textbf{Dummy Code} & 
\textbf{\makecell{Pearson \\ Correlation (Mean)}} & 
\textbf{\makecell{ANOVA \\ F-Statistic (p $< 0.001$)}} & 
\textbf{Granger Causality} & 
\textbf{\makecell{Mean Elapsed \\ Time Increase}} & 
\textbf{\makecell{Power Consumption \\ Behavior}} \\
\midrule
\makecell[l]{One For Loop} & 
0.647 & 
19.83 & 
\makecell[l]{Strong in moderate\\ runs} & 
Moderate increase& 
\makecell[l]{More consistent but\\ minor spikes} \\
\makecell[l]{Two Nested For Loops} & 
0.848 & 
72.42 & 
\makecell[l]{Strong in most runs} & 
\makecell[l]{Significant \\ increase} & 
\makecell[l]{Higher spikes and larger\\variability} \\
\makecell[l]{Single Function} & 
0.459 & 
62.06 & 
\makecell[l]{Fewer significant \\ lags overall} & 
\makecell[l]{Small increase} & 
\makecell[l]{Flatter pattern with few\\ power spikes} \\
\makecell[l]{If Statement} & 
0.415 & 
196.68 & 
\makecell[l]{Causality in specific runs} & 
\makecell[l]{Minimal \\ increase} & 
\makecell[l]{Conditional jumps cause \\ mild spikes} \\
\bottomrule
\end{tabular}
\end{adjustbox}
\end{table*}

\subsubsection{Dummy Code Injection Methodology}
\label{sec:dummy_injection}

To subtly disrupt Mirai’s scanning-phase power signatures~\cite{ning2022trojanflow}, we inject controlled dummy operations into its source code. The goal is to perturb high-impact temporal segments of the power trace without impairing the malware’s core functionality, thereby deceiving machine learning-based detection systems~\cite{lipp2021platypus}. Each dummy code variant is designed to target different aspects of computational overhead and execution variability, enabling a spectrum of perturbation intensities.

\begin{itemize} [leftmargin=*]
  \item \textbf{Single Function:} Implements lightweight arithmetic or logical operations within a dedicated function, introducing minimal computational load. This variant produces smooth, low-amplitude fluctuations, making it effective for stealthy perturbations where subtlety is critical.
  \item \textbf{One For Loop:} Adds a single bounded iteration loop performing repetitive operations. This design yields consistent, moderate power spikes that disrupt temporal regularity while maintaining low execution overhead.
  \item \textbf{Two Nested Loops:} Incorporates nested iterations to significantly amplify variability in power consumption. Although more disruptive to detection models, this variant introduces higher execution latency, making it a trade-off between evasion strength and runtime cost.
  \item \textbf{If Statement:} Inserts conditional branches that trigger selective operations based on lightweight, pseudo-random criteria. The resulting irregular, non-periodic spikes can break the periodicity patterns often exploited by sequential detection models.
\end{itemize}

All dummy code is strategically injected into Mirai’s scanning logic (Figure~\ref{fig:AttackStructure} - Client Side), ensuring perturbations occur during the most classifier-sensitive phase identified by our SHAP-based feature importance analysis. Each variant is implemented with computational constraints of smartphone-class IoT devices in mind, avoiding excessive memory or CPU usage that could cause execution failures. The resulting perturbed malware samples form the basis for our cross-architecture evaluation, enabling a controlled study of how different perturbation patterns affect detection robustness and power signature profiles.

\subsubsection{Power Analysis and Detection Metrics}
\label{sec:power_analysis}

To rigorously evaluate the effects of dummy code injection on detection reliability, we employ a combination of statistical analyses and runtime profiling methods. These techniques quantify both the magnitude and nature of power trace perturbations, allowing us to link observed changes directly to classifier performance degradation.

\begin{itemize} [leftmargin=*]
    \item \textbf{Pearson Correlation Coefficients:} Measure the linear relationship between dummy code execution and instantaneous power fluctuations~\cite{pearson_correlation}. High correlation values indicate that injected operations consistently affect the power profile, while lower values suggest more diffuse, less predictable impacts.
    \item \textbf{ANOVA (Analysis of Variance):} Test the statistical significance of mean differences in power consumption across dummy code variants~\cite{anova_statistical}. This highlights which injection patterns produce the most distinct deviations in average power draw relative to the baseline Mirai scanning phase.
    \item \textbf{Granger Causality Tests:} Determine whether dummy code execution can be said to causally influence future power measurements~\cite{granger_causality}. Strong causality scores, especially for high-complexity variants such as nested loops, reinforce the interpretation that these injections directly manipulate side-channel signals.
    \item \textbf{Dynamic Profiling:} Record fine-grained runtime timestamps to align dummy code execution intervals with observable power deviations~\cite{dynamic_profiling}. This provides temporal context, confirming that power anomalies occur synchronously with the injected operations.
\end{itemize}

By combining these methods, we can characterize how each dummy code variant alters the underlying side-channel signal—both in statistical structure and in temporal dynamics. This multi-faceted analysis allows us to directly connect perturbation patterns to changes in model detection confidence, helping to identify which injection strategies most effectively degrade ML-based side-channel detectors without compromising malware functionality.

\subsubsection{Impact on Power Patterns}
\label{sec:impact_patterns}

The introduction of dummy code variants alters Mirai’s power side-channel behavior through a combination of increased computational activity and timing irregularities. First, dummy code raises the overall computational workload, leading to higher power amplitude and more pronounced spikes in the consumption trace~\cite{power_amplitude_increase}. Second, the injected code introduces execution delays, which reduce the frequency of scanning operations and distort the periodic patterns typically associated with Mirai’s behavior~\cite{frequency_reduction}. Finally, these combined effects enhance the stealthiness of the malware by deviating from its baseline power signature, thereby increasing the likelihood of misclassification by anomaly detection models~\cite{stealthy_attacks}.

Experimental results summarized in Table~\ref{table:DummyCodeAnalysis} reveal variant-specific trade-offs between stealthiness, computational overhead, and disruption effectiveness. The ``One For Loop'' variant emerges as the optimal balance between evasion capability and operational efficiency, while ``Two Nested Loops'' achieves stronger perturbation but introduces noticeable timing artifacts. All ANOVA tests achieved statistical significance ($p < 0.001$ for all variants), confirming that dummy code injections caused measurable differences in power signatures.
Each reported metric represents the mean value across 200 independent runs per variant; per-run variability was low and omitted from the table for brevity.

These findings demonstrate the practical feasibility of adversarial dummy code injection for undermining power-based malware detection and highlight critical vulnerabilities requiring adaptive, robust defense strategies.

\section{Performance Evaluation}
\label{sec:performance_evaluation}

This section evaluates the effectiveness of the proposed adversarial dummy code injection strategy against AI/ML-based power side-channel malware detection systems. We first describe the experimental setup and dataset collection process, followed by an analysis of baseline power signatures. We then assess the impact of dummy code perturbations on power patterns and classifier performance, concluding with a discussion of attack success rates and ethical considerations.

\subsection{Experimental Setup}
\label{sec:experimental_setup}

\begin{figure}[ht]
\centerline{\includegraphics[width=8cm]{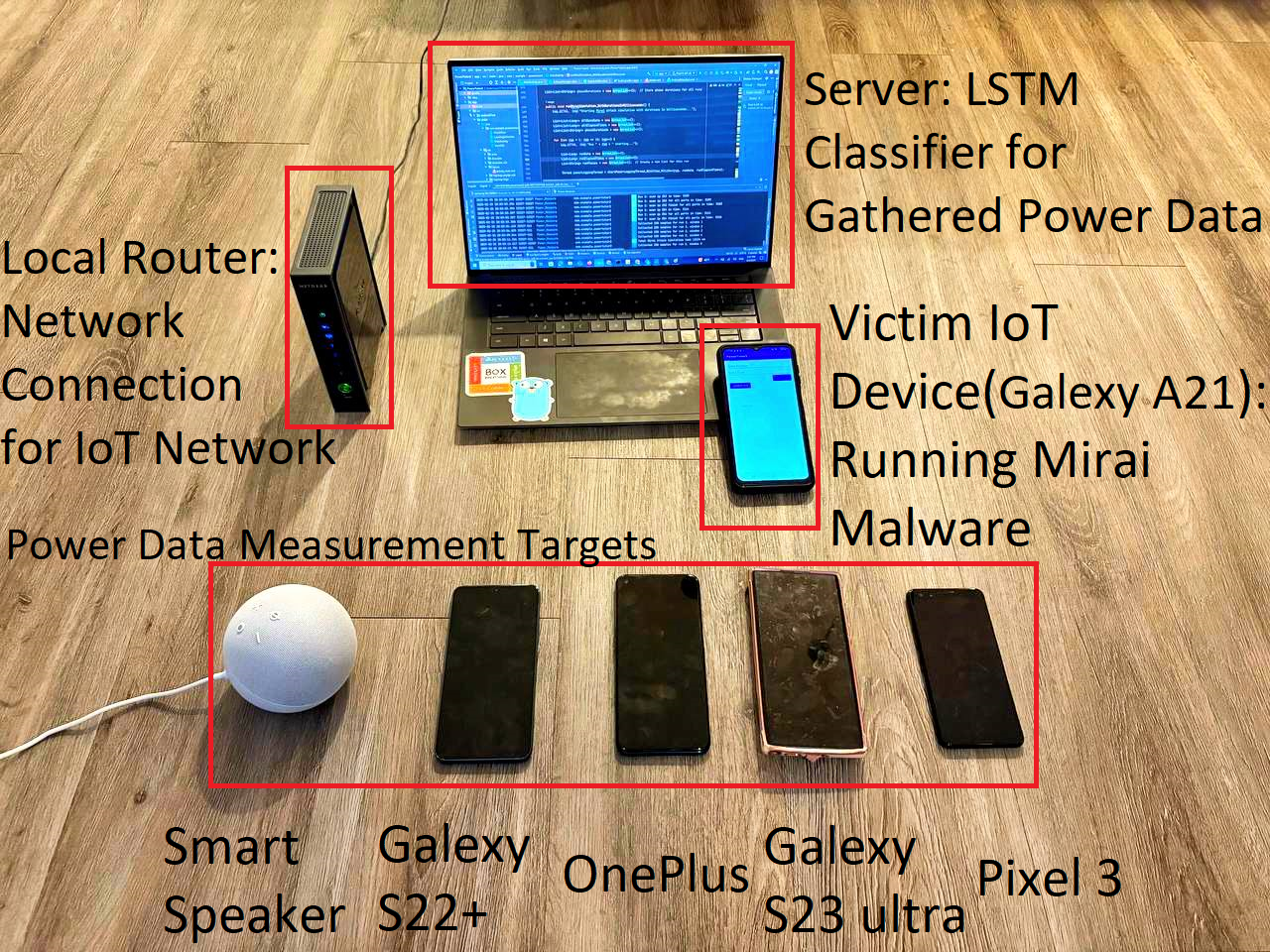}}
\caption{Experimental setup: IoT device running Mirai variant, server hosting the detection model, and local router for isolated communication.}
\label{fig:realpic}
\end{figure}

\textbf{IoT device configuration.}  
Experiments were conducted using five Android smartphones with diverse hardware and power management architectures: Samsung Galaxy S23 Ultra, Galaxy S22+, Galaxy A21, Google Pixel 3, and OnePlus. This diversity enhances generalizability across varying device characteristics, consistent with prior smartphone-based side-channel research~\cite{oberhuber2025power,muhammad2023smartphone}. Devices were configured to operate within a local IoT network alongside an Alexa device and additional smartphones, all communicating through a private router without internet exposure (Figure~\ref{fig:realpic}). While this evaluation focuses on smartphone-class IoT devices, future testing on embedded Linux and RTOS-based IoT platforms is necessary to assess broader applicability.

\textbf{Dataset collection and processing.}  
Existing datasets such as CICIDS2017~\cite{sharafaldin2018toward} and BotIoT~\cite{koroniotis2019towards} lack power consumption data required for this study. Therefore, we generated a custom power-side channel dataset capturing three operational states: Idle, IoT Service, and Mirai Botnet execution.

Power measurements were obtained using the Android Power Management Library~\cite{android_power_management}, which accesses device fuel gauge chips such as the Summit SMB347 and Maxim MAX17050. Key metrics include instantaneous current, remaining charge, and battery capacity, with primary analysis based on \texttt{BATTERY\_PROPERTY\_CURRENT\_NOW} property, capturing real-time battery current in microamperes at a sampling rate of 1 kHz.

For each IoT device, we collected 340 labeled runs per class, resulting in 1020 samples per device. Across five devices, the final dataset contains 5100 labeled records. Each record consists of a window of 150 consecutive power measurements, preserving temporal characteristics. At a sampling rate of 1kHz, each window corresponds to a 150-millisecond segment of continuous power measurements, capturing short-term temporal dynamics crucial for classifying malware behavior. The dataset was partitioned into 85\% training and 15\% testing/validation sets. This diverse, structured dataset enables robust evaluation of anomaly detection models across different device architectures and usage scenarios.

The dataset will be made publicly available after publication to support reproducibility and further research.

\textbf{Mirai botnet modification.}  
We customized Mirai using its open-source GitHub repository~\cite{mirai_github}, injecting structured dummy code to perturb its power signature while preserving functional malware behavior. Power traces were segmented into fixed-size windows aligned with model input requirements.

\textbf{Evaluated classifiers and preprocessing pipelines.}  
To assess the generalization and robustness of power-based malware detection, we trained and evaluated six deep learning architectures on the collected power traces. These include Long Short-Term Memory (LSTM) networks, which capture sequential dependencies in time-series power data~\cite{jin2024pca-lstm}, and Bidirectional LSTMs (BiLSTM), which improve anomaly classification by incorporating both past and future context~\cite{yan2024bilstm-cnn}. We also evaluate Temporal Convolutional Networks (TCN), which apply dilated 1D convolutions to model long-range temporal patterns in power traces~\cite{li2023tcn-energy}, and a hybrid BiLSTM+CNN model that leverages convolutional layers for local feature extraction followed by BiLSTM-based sequence modeling to enhance resilience to localized perturbations~\cite{wang2022fiber-fault}. Additionally, we include a reconstruction-based architecture that combines LSTM Autoencoders with a Multi-Layer Perceptron (MLP) classifier, enabling the model to learn benign behavior patterns and flag deviations as anomalies~\cite{smartmeter2021bilstm}.

\begin{table}[ht]
\centering
\caption{Classification performance on clean traces (no adversarial perturbations). Best results in bold.}
\label{table:classification_report}
\resizebox{\columnwidth}{!}{%
\begin{tabular}{lcccc}
\toprule
\textbf{Model} & \textbf{Accuracy} & \textbf{Precision} & \textbf{Recall} & \textbf{F1-score} \\
\midrule
LSTM & 0.64 & 0.65 & 0.64 & 0.63 \\
BiLSTM & 0.91 & 0.90 & 0.90 & 0.89 \\
TCN & \textbf{0.95} & \textbf{0.95} & 0.94 & \textbf{0.94} \\
BiLSTM+CNN & \textbf{0.96} & 0.94 & \textbf{0.96} & \textbf{0.95} \\
CNN+Attention & 0.72 & 0.73 & 0.72 & 0.72 \\
LSTM+AE+MLP & 0.65 & 0.67 & 0.67 & 0.65 \\
\bottomrule
\end{tabular}
}
\end{table}

\begin{figure}[ht]
    \centering
    \includegraphics[width=0.48\textwidth]{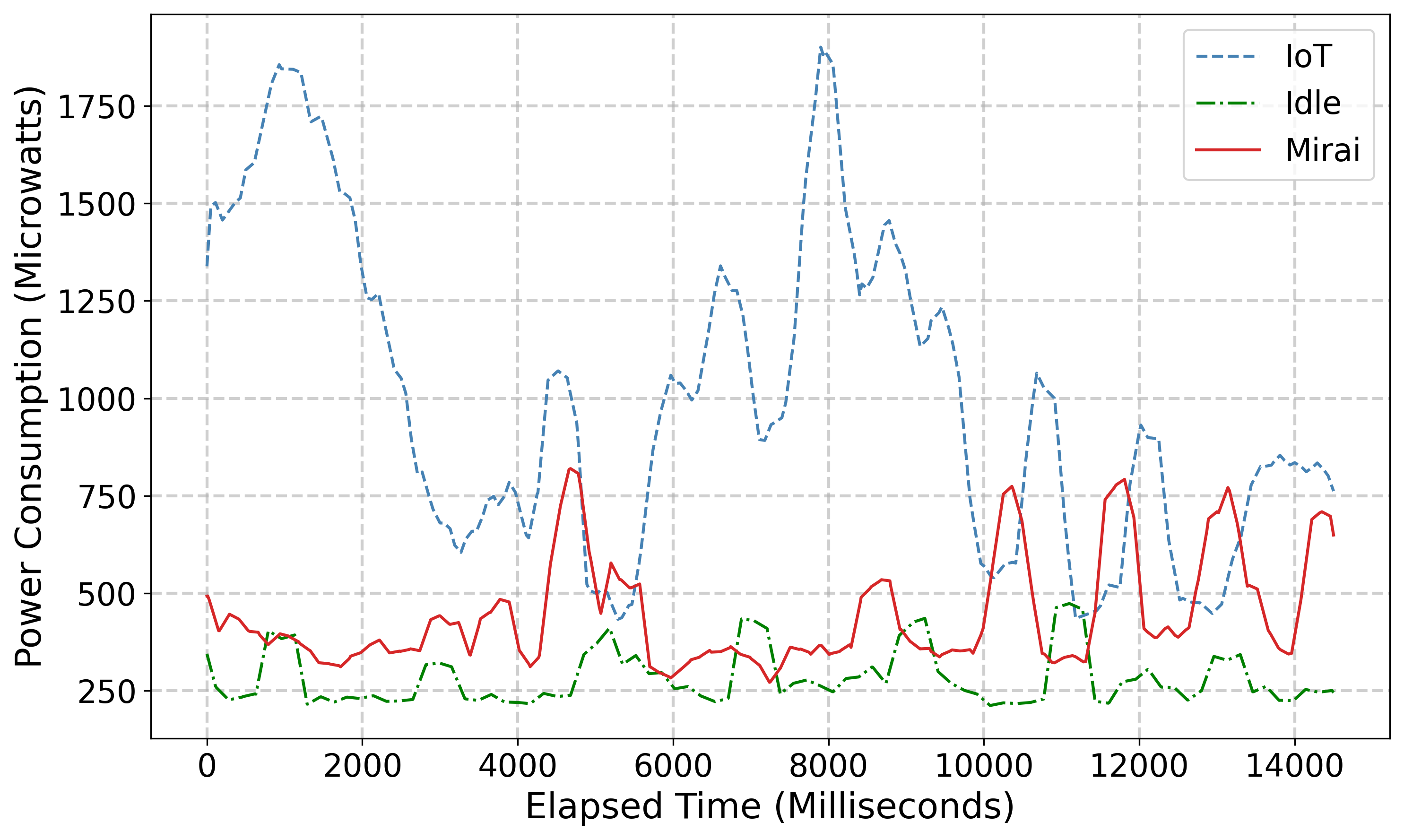
    }
   \caption{Comparison of power consumption across IoT operational states. The figure shows Mirai’s power trace alongside IoT service and Idle states, highlighting deviations caused by scanning activity compared to benign behavior.}

    \label{fig:power_comparison}
\end{figure}

Preprocessing included autocorrelation analysis for periodicity detection~\cite{jin2022enhancing}, wavelet transforms for capturing transient anomalies~\cite{bae2023autoscaled}, and principal component analysis (PCA) for dimensionality reduction~\cite{soni2022low}. All features were standardized using \texttt{StandardScaler}. Hyperparameters such as learning rate, batch size, and early stopping patience were optimized via grid search~\cite{budiarso2024optimizing}. Models were trained using 4-fold cross-validation over 50 epochs with early stopping, and evaluated locally on the edge device (Raspberry Pi 4B, 4GB RAM, quad-core Cortex-A72).

Table~\ref{table:classification_report} summarizes baseline classification performance on clean power traces (i.e., without adversarial dummy code injection), reporting accuracy, precision, recall, and F1-score averaged across Idle, IoT Service, and Mirai classes. Among the evaluated models, \textbf{BiLSTM+CNN} and \textbf{TCN} achieved the highest accuracy. The TCN’s use of dilated convolutions enables the detection of long-range temporal patterns, while BiLSTM+CNN leverages both local filtering and global sequence modeling, improving robustness to noise and hardware variability. CNN+Attention also demonstrated moderate performance, benefiting from its ability to emphasize informative power trace segments.

\begin{figure}[ht]
    \centering
    \begin{subfigure}[b]{0.48\textwidth}
        \centering
        \includegraphics[width=\textwidth]{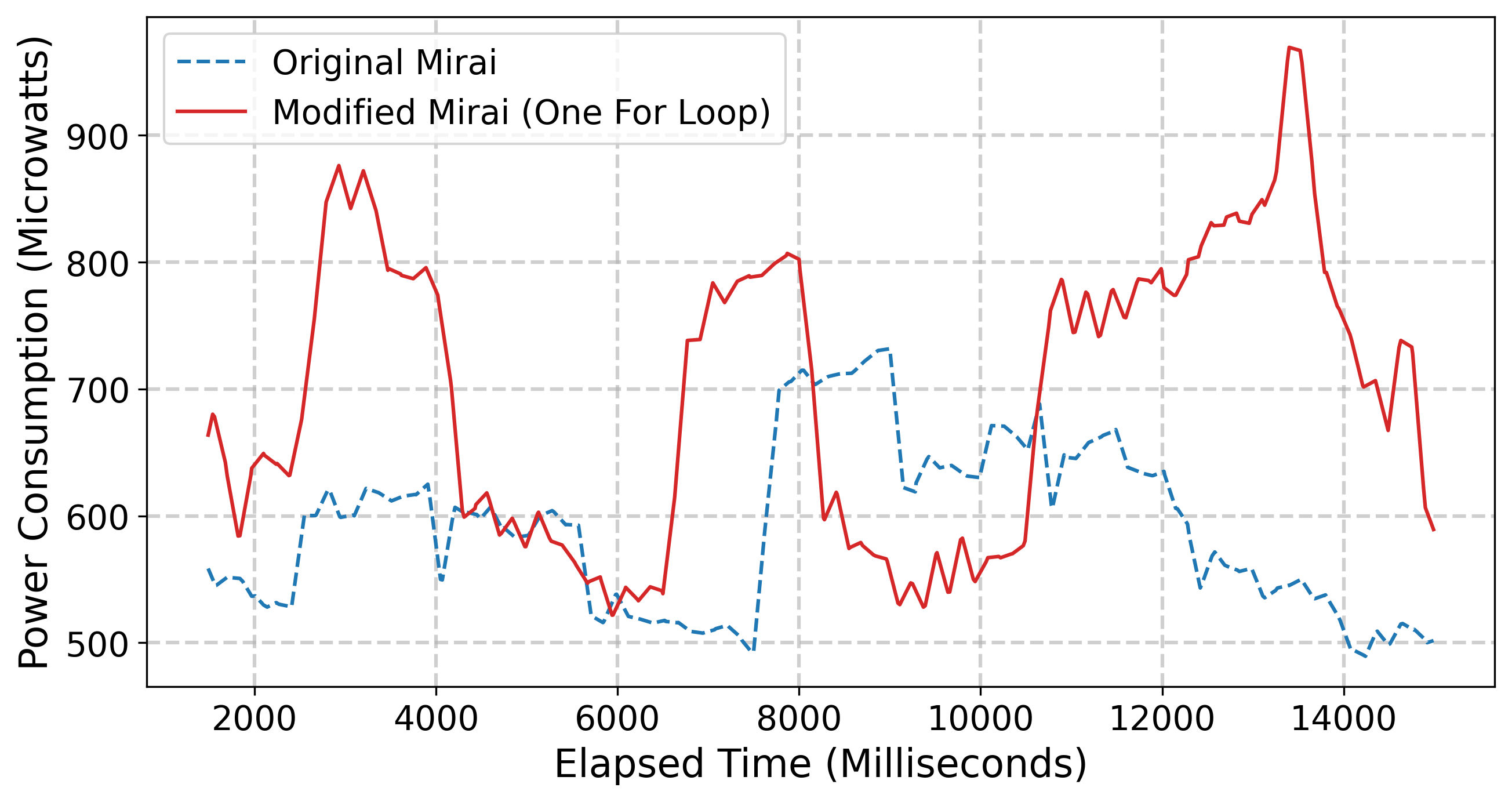}
        \caption{Power consumption: Original vs. Modified Mirai (One For Loop).}
        \label{fig:mirai_modified_power_one_forloop}
    \end{subfigure}
    \hfill
    \begin{subfigure}[b]{0.48\textwidth}
        \centering
        \includegraphics[width=\textwidth]{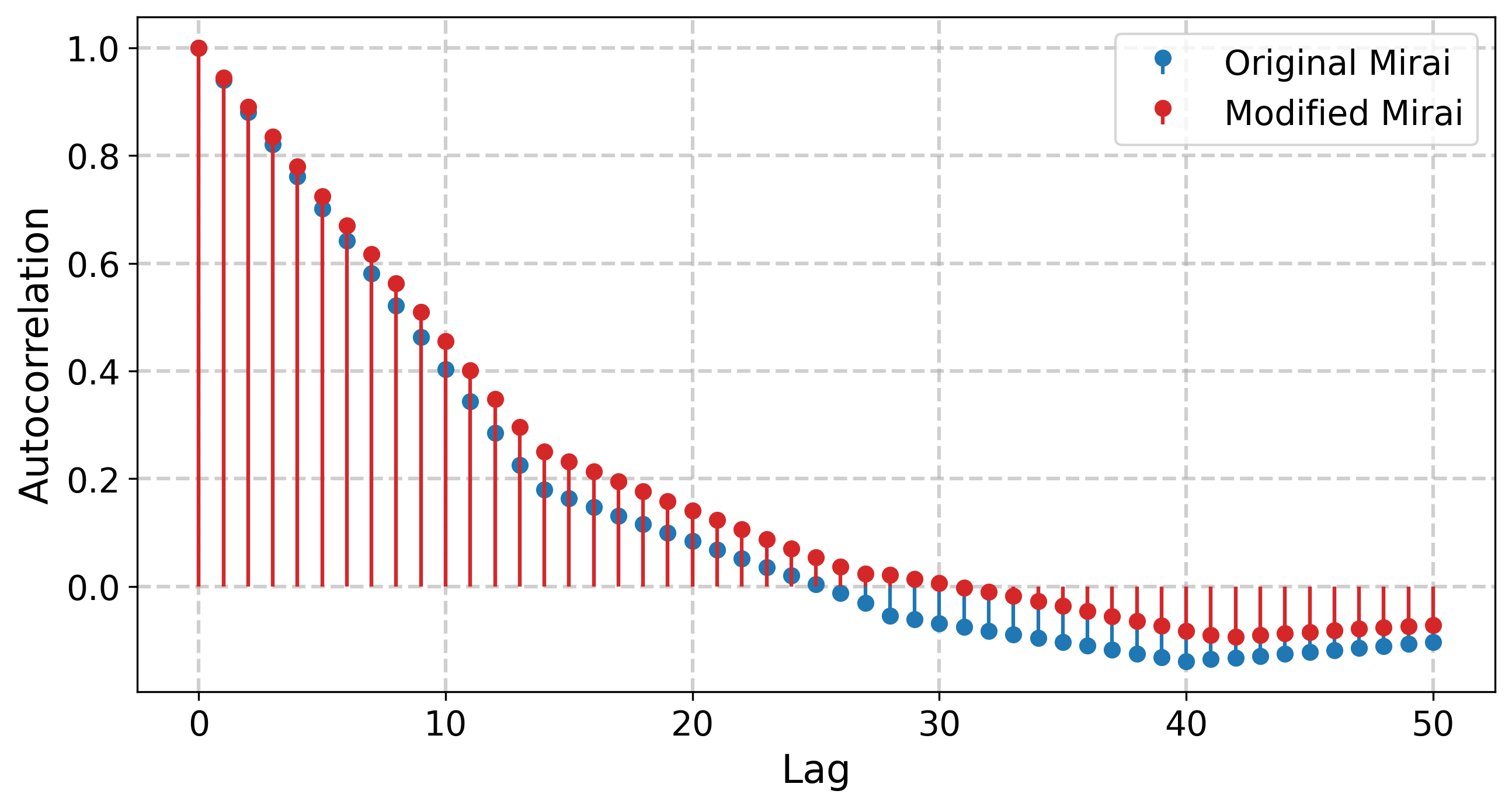}
        \caption{Autocorrelation: Original vs. Modified Mirai (One For Loop).}
        \label{fig:mirai_modified_autocorrelation_one_forloop}
    \end{subfigure}
    \caption{Impact of single for-loop dummy code injection on Mirai’s power signature and periodicity.}
    \label{fig:mirai_vs_modified_one_forloop}
\end{figure}

\begin{figure}[ht]
    \centering
    \begin{subfigure}[b]{0.48\textwidth}
        \centering
        \includegraphics[width=\textwidth]{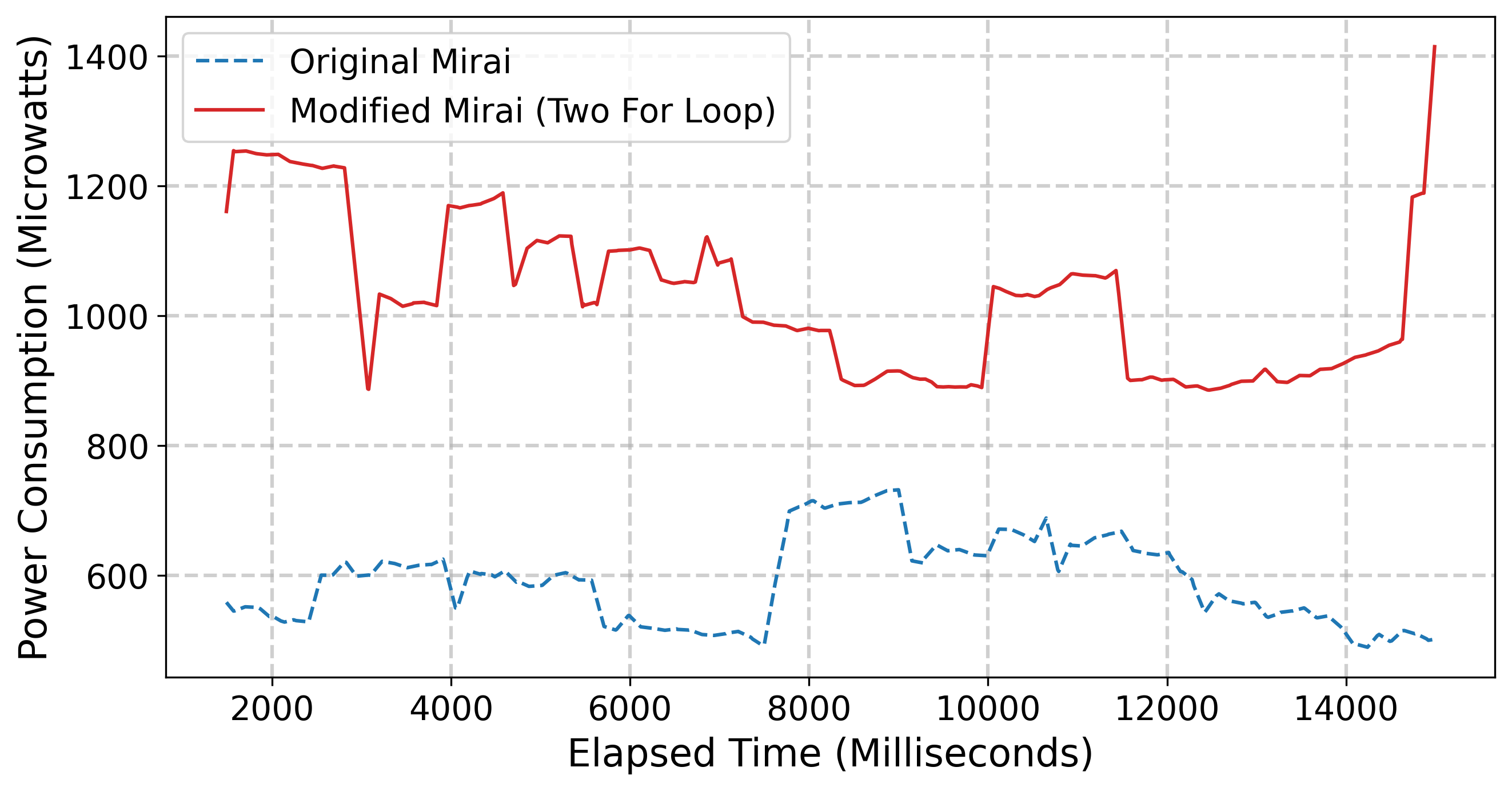}
        \caption{Power consumption: Original vs. Modified Mirai (Two Nested Loops).}
        \label{fig:mirai_modified_power_two_forloop}
    \end{subfigure}
    \hfill
    \begin{subfigure}[b]{0.48\textwidth}
        \centering
        \includegraphics[width=\textwidth]{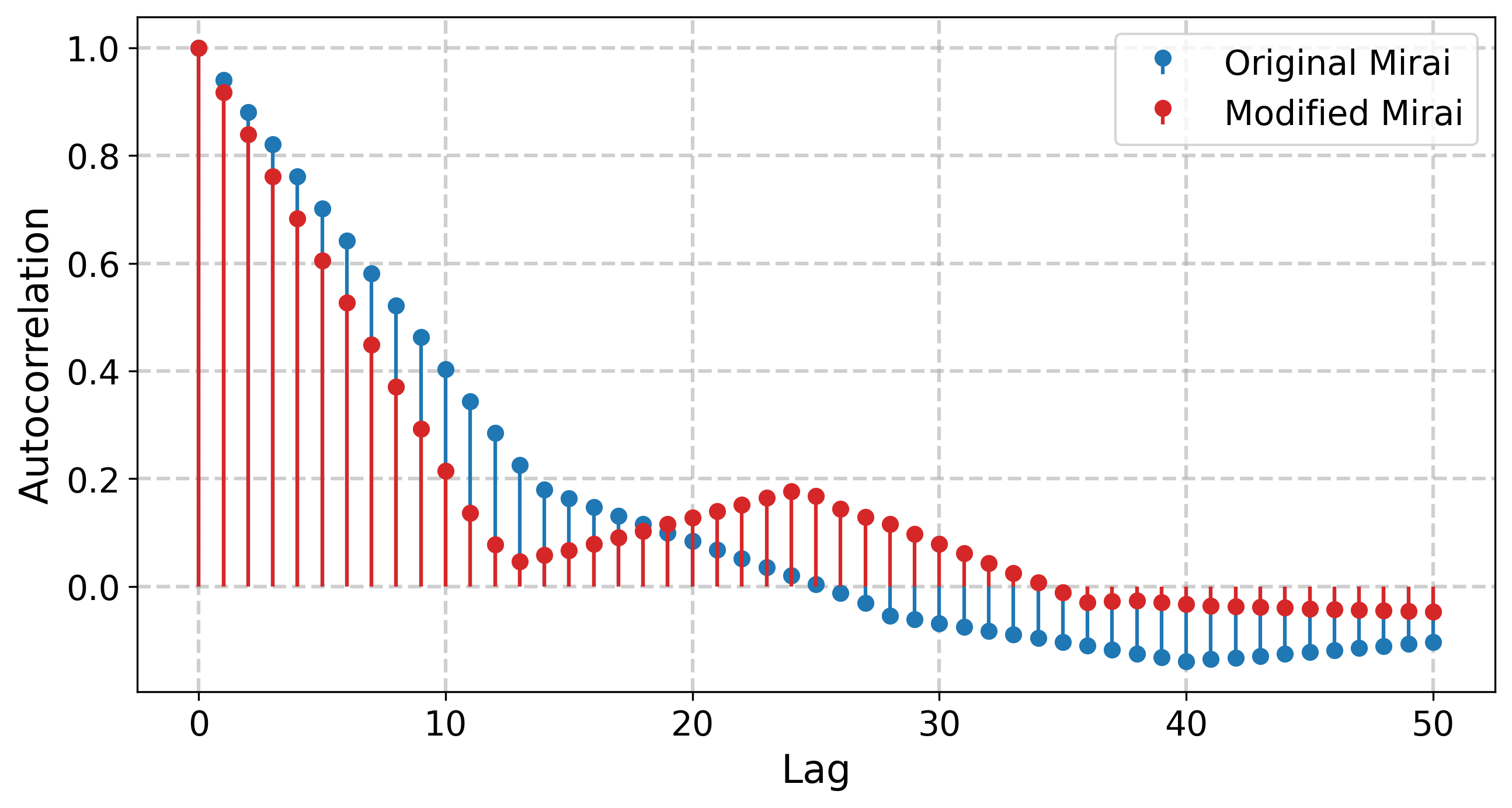}
        \caption{Autocorrelation: Original vs. Modified Mirai (Two Nested Loops).}
        \label{fig:mirai_modified_autocorrelation_two_forloop}
    \end{subfigure}
    \caption{Impact of two nested loops dummy code injection on Mirai’s power signature and periodicity.}
    \label{fig:mirai_vs_modified_two_forloop}
\end{figure}

\subsection{Baseline Power Signature Analysis}
\label{sec:baseline_power_analysis}

Before introducing adversarial perturbations, we analyze the baseline power consumption patterns of IoT devices under three states: Idle, IoT Service Execution, and Mirai Botnet activity. This comparison provides foundational insights into how different behaviors manifest in power side-channels and reveals opportunities for evasion.

As shown in Figure~\ref{fig:power_comparison}, the idle state exhibits a flat and stable power profile, with only minor fluctuations caused by background processes. IoT service execution introduces moderate variability, characterized by intermittent bursts corresponding to event-driven tasks such as sensing and communication. In contrast, Mirai botnet activity produces a highly structured and distinctive power signature, marked by frequent, high-amplitude spikes during its scanning phase. These spikes result from intensive IP and port scanning operations and create strong periodicity in the power trace.

Overall, the baseline analysis shows that benign activities generate irregular, low-amplitude patterns, while Mirai introduces regular, high-magnitude anomalies. This contrast enables machine learning-based anomaly detectors to distinguish malicious behavior, but it also exposes a vulnerability: the predictability of Mirai’s signature makes it possible to conceal it within normal variability using carefully designed runtime perturbations, motivating the adversarial strategies explored in the following sections.

\subsection{Impact of Dummy Code Injection}
\label{sec:dummy_code_impact}

This section analyzes how dummy code injections perturb Mirai's power signatures and evaluates the statistical significance, causality, and operational trade-offs introduced by different dummy code variants.

\subsubsection{Power Pattern Perturbations}
\label{sec:pattern_perturbations}

Injecting structured dummy code into Mirai’s scanning phase visibly changes the timing and intensity of its power side-channel signals. Figures~\ref{fig:mirai_vs_modified_one_forloop} and \ref{fig:mirai_vs_modified_two_forloop} compare the smoothed power traces and autocorrelation curves for the original Mirai against its modified versions.

A single for-loop variant adds lightweight computational overhead, introducing small phase shifts and moderate amplitude fluctuations (Figure~\ref{fig:mirai_modified_power_one_forloop}). These alterations blur the strong periodic patterns that detection models often rely on. In contrast, the two nested loops variant produces far greater disruption, with frequent, irregular power spikes that significantly weaken autocorrelation strength (Figure~\ref{fig:mirai_modified_power_two_forloop}). This makes it harder for sequential models to match the signal to a known malware profile.

\begin{figure}[ht]
    \centering
    \begin{subfigure}[b]{0.48\textwidth}
        \centering
        \includegraphics[width=\textwidth]{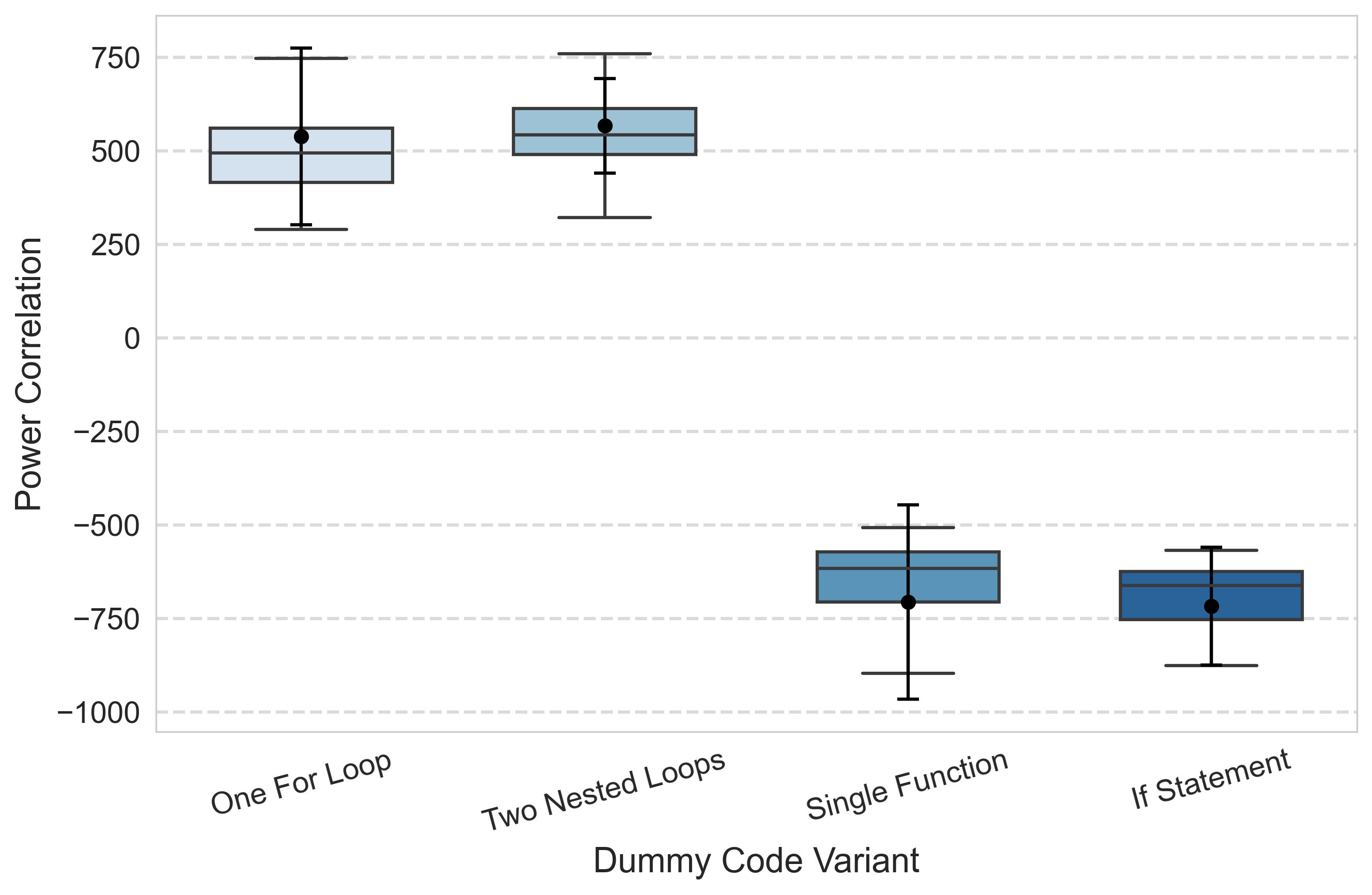}
        \caption{ANOVA Box Plot.}
        \label{fig:anova_boxplot}
    \end{subfigure}
    \hfill
    \begin{subfigure}[b]{0.48\textwidth}
        \centering
        \includegraphics[width=\textwidth]{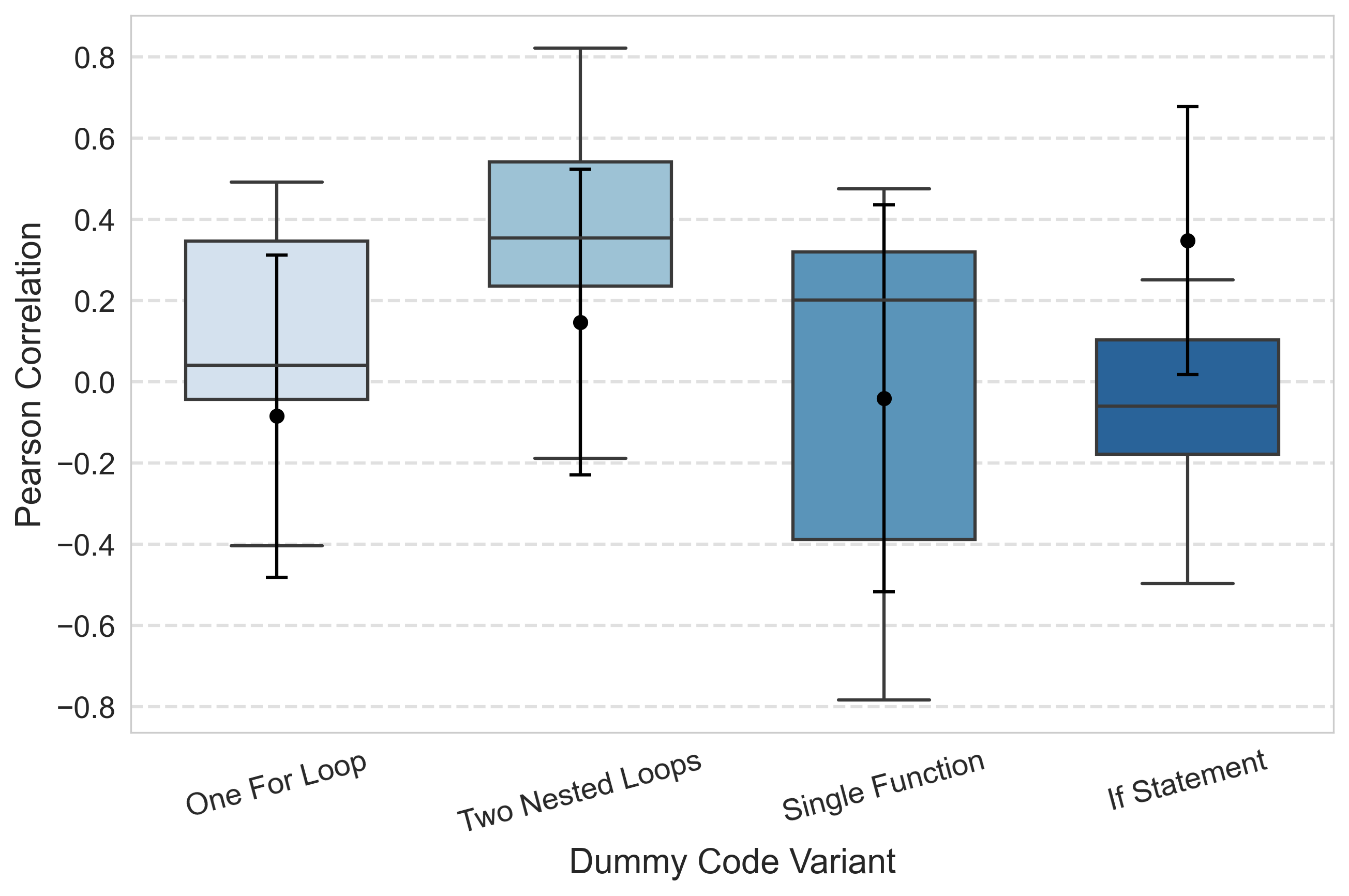}
        \caption{Pearson Correlation Box Plot.}
        \label{fig:pearson_boxplot}
    \end{subfigure}
    \caption{Statistical comparison of original versus modified Mirai using ANOVA F-statistics and Pearson correlation.}
    \label{fig:anova_pearson}
\end{figure}
\subsubsection{Statistical Analysis of Perturbations}
\label{sec:statistical_impact}

To move beyond visual inspection, we quantified the effect of each dummy code variant using statistical methods:

\begin{itemize}[leftmargin=*]
    \item \textbf{Pearson Correlation Analysis:}  
    Correlation scores between original and modified traces measure similarity in power patterns. Lower values mean stronger perturbations. As shown in Figure~\ref{fig:pearson_boxplot}, nested loops consistently achieve the lowest correlations, showing that their signal profile deviates the most from the baseline.
    
    \item \textbf{ANOVA (Analysis of Variance):}  
    ANOVA testing confirms that the power distributions of modified variants are statistically distinct from the original Mirai traces (Figure~\ref{fig:anova_boxplot}). High F-statistics, coupled with $p < 0.001$ for all cases, indicate substantial divergence in mean power levels across variants. This statistically verifies that each dummy code strategy creates a unique, measurable shift in side-channel behavior.
\end{itemize}

\subsubsection{Causality Verification}
\label{sec:causality_analysis}

While correlation analysis can reveal statistical associations, it does not prove that dummy code execution directly drives changes in side-channel power profiles. To establish a stronger causal link, we conducted Granger causality tests across all dummy code variants and multiple temporal lags. Granger causality assesses whether past values of one time series (dummy code activity) provide statistically significant predictive power for future values of another time series (power consumption), beyond what the target series can predict from its own history.

\begin{figure}[ht]
    \centering
    \includegraphics[width=0.5\textwidth]{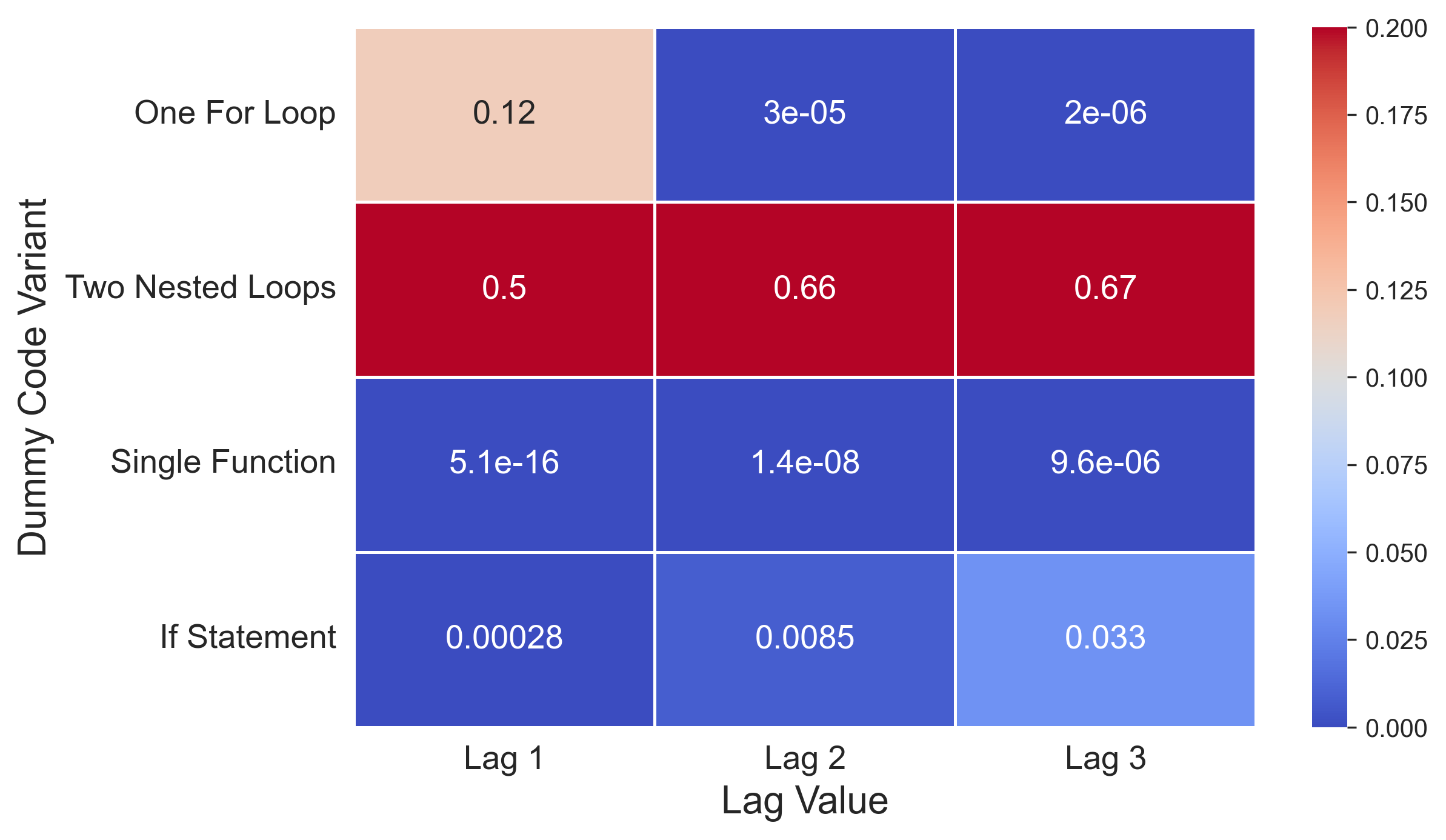}
    \caption{Granger causality heatmap illustrating the causal impact of dummy code variants on power fluctuations.}
    \label{fig:granger_heatmap}
\end{figure}

The results, shown in Figure~\ref{fig:granger_heatmap}, confirm that variants with more computationally intensive structures exert a stronger causal influence on power fluctuations. For example, the ``Two Nested Loops'' variant consistently yields high causality scores across all three lag values, with significance levels reaching 0.5, 0.66, and 0.67 for Lag 1 through Lag 3, respectively. This persistent influence suggests that repeated and nested operations generate sustained side-channel effects that remain detectable over multiple sampling intervals. In contrast, lightweight perturbations such as the ``If Statement'' variants produce comparatively low causality values, particularly at longer lags, indicating a more transient or localized effect on power traces. 

These findings strengthen the argument that dummy code can be precisely tuned to control side-channel signatures, with complexity directly translating to stronger and more persistent causal impact. Moreover, by validating the perturbation effect through causality analysis rather than correlation alone, we provide more definitive evidence that the observed power variations are an intentional byproduct of adversarial manipulation rather than incidental system noise. This causal confirmation is critical for both characterizing the attack’s potency and informing the design of future side-channel defenses that account for temporally sustained adversarial effects.

\subsubsection{Execution Time Impact and Trade-offs}
\label{sec:execution_time_impact}

\begin{figure}[h]
    \centering
    \begin{subfigure}[b]{0.48\textwidth}
        \centering
        \includegraphics[width=\linewidth]{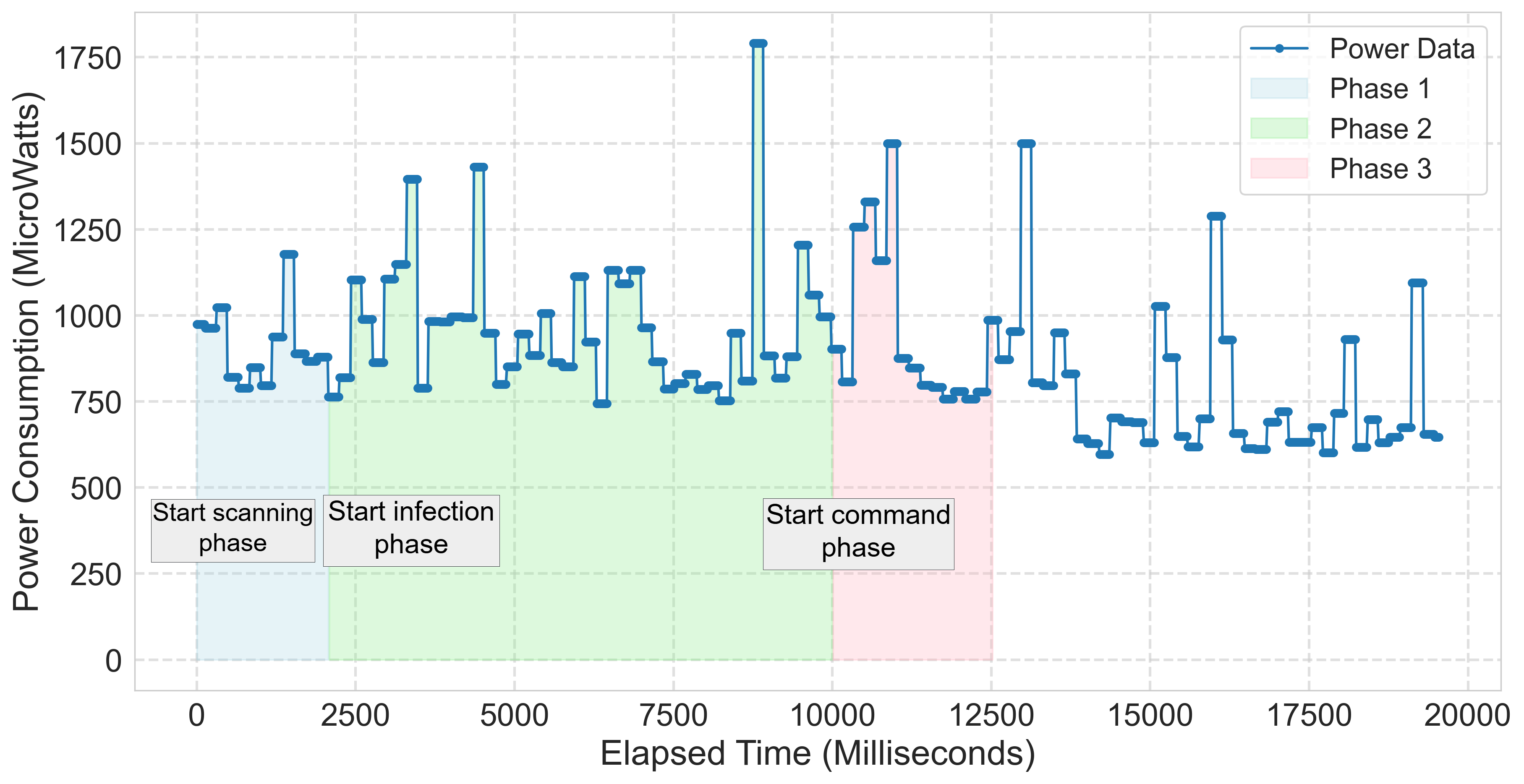}
        \caption{Elapsed Time: Original Mirai.}
        \label{fig:mirai_elapsed}
    \end{subfigure}
    \hfill
    \begin{subfigure}[b]{0.48\textwidth}
        \centering
        \includegraphics[width=\linewidth]{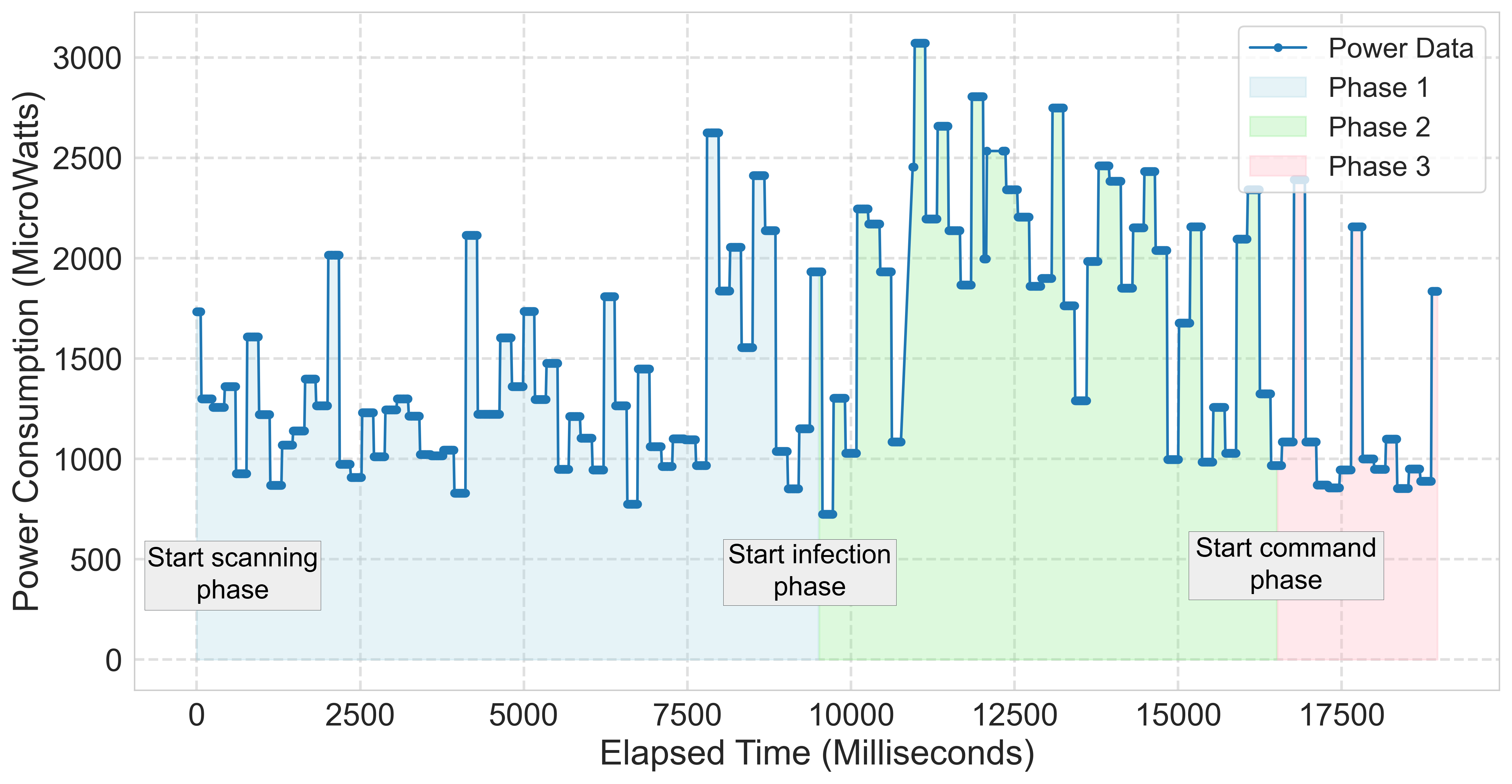}
        \caption{Elapsed Time: Modified Mirai (Dummy Code Injected).}
        \label{fig:modified_mirai_elapsed}
    \end{subfigure}
    \caption{Comparison of execution time between original and dummy-code-injected Mirai variants.}
    \label{fig:time_comparison}
\end{figure}

While dummy code variants effectively disrupt power patterns, they also introduce execution overhead. Figure~\ref{fig:time_comparison} compares elapsed time for original and modified Mirai executions. Variants such as two nested loops significantly prolong scanning phase durations, potentially increasing detectability through timing analysis. 
Nevertheless, even the highest-overhead variant remained within the computational capacity of smartphone-class IoT devices, confirming operational feasibility without causing execution failure or resource exhaustion. This measurable execution overhead suggests a potential detection strategy: defenders could baseline typical scanning phase durations and flag anomalous increases introduced by dummy code injection. While our attack remains stealthy in power signatures, it incurs timing penalties that may serve as an auxiliary detection signal in multi-modal monitoring systems.

\begin{table*}[ht]
\caption{Summary of optimal dummy code variants across attack objectives.}
\label{table:DummyCodeSummary}
\centering
\resizebox{\textwidth}{!}{%
\begin{tabular}{lll}
\toprule
\textbf{Aspect} & \textbf{Best Dummy Code} & \textbf{Reason} \\
\midrule
Stealthiness (minimal power change) & Single Function & It has the lowest correlation, smoother power consumption, and fewer spikes. \\
Disruption (power pattern variability) & Two Nested For Loops & Consistent Granger causality across runs indicate maximum variability. \\
Time Efficiency & If Statement & Minimal elapsed time increase while still introducing noticeable changes. \\
Balanced Approach & One For Loop & Balances power pattern variability and execution time reasonably well. \\
\bottomrule
\end{tabular}%
}
\end{table*}

\subsection{Adversarial Attack Effectiveness}
\label{sec:attack_effectiveness}

This section quantifies the effectiveness of dummy code perturbations in evading AI/ML-based side-channel detection systems. We report Attack Success Rates (ASR) across different dummy code variants, analyze misclassification trends, and discuss trade-offs between stealthiness, disruption, and operational efficiency.

\subsubsection{Attack Success Rate (ASR) Evaluation}
\label{sec:attack_success_rate}
ASR quantifies the proportion of Mirai samples that are misclassified as benign (Idle or IoT Service) following the injection of dummy code. Table~\ref{table:asr_baseline} reports mean ASR values averaged over 200 independent runs per dummy variant, ensuring consistent evaluation across trials.

Among the tested variants, the one for loop and two nested loops achieved the highest ASRs, reaching 91.28\% and 96.4\% respectively under LSTM-based models. Overall, LSTM and LSTM-AE models exhibited the greatest vulnerability across all variants, suggesting that sequential architectures are particularly sensitive to timing-based perturbations introduced by adversarial logic.

In contrast, CNN+Attention and BiLSTM models showed stronger resistance. CNN+Attention achieved the lowest ASR (42\%) under the if-statement attack, while BiLSTM remained under 78\% in most cases. Temporal Convolutional Networks (TCNs) demonstrated moderate robustness, maintaining ASR below 68\% across the board.

These results highlight how architectural design influences evasion susceptibility. Despite analyzing power signals rather than visual inputs, power-based ML detectors are comparably vulnerable to adversarial evasion, with attack success rates similar to those seen in vision domains~\cite{athalye2018obfuscated, carlini2017adversarial}, underscoring the need for adversarially robust side-channel defenses.

\begin{table*}[ht]
\centering
\small % or \footnotesize to make it even smaller
\caption{Attack Success Rate (ASR\%) for baseline models across dummy code variants. Higher ASR indicates greater vulnerability.}
\label{table:asr_baseline}
\begin{tabular}{lcccccc}
\toprule
\shortstack{\textbf{Dummy Code} \\ \textbf{Variant}} & 
\textbf{LSTM} & 
\textbf{BiLSTM} & 
\textbf{TCN} & 
\shortstack{\textbf{BiLSTM +CNN}} & 
\shortstack{\textbf{CNN+Attn}} & 
\shortstack{\textbf{LSTM+AE}} \\
\midrule
Single Function     & 90.0  & 50.8 & 62.1 & 50.0 & 47.3 & 76.2 \\
One For Loop        & 91.2  & 67.0 & 67.5 & 71.2 & 61.6 & 83.7 \\
Two For Loops       & 96.4  & 57.2 & 61.7 & 72.0 & 45.3 & 89.0 \\
If Statement        & 83.3  & 77.3 & 59.6 & 77.3 & 42.0 & 79.0 \\
\bottomrule
\end{tabular}
\end{table*}

\subsubsection{Misclassification Trends Across Dummy Code Variants}
\label{sec:misclassification_trends}

Misclassification trends highlight trade-offs between stealth, execution overhead, and evasion success, as shown in Table~\ref{table:DummyCodeSummary}. Stealth-focused variants like the single function and if statement introduce minimal power deviations and achieve modest ASRs with low timing impact, ideal when stealth is prioritized. In contrast, the two nested loops variant maximizes ASR by heavily distorting power patterns but incurs high overhead, making it vulnerable to timing-aware defenses. The one for loop variant offers a balanced approach, achieving effective evasion with moderate resource use.

These results demonstrate that carefully crafted runtime modifications can evade power-based malware detection. Dummy code enables tunable control over stealth and disruption, allowing adversaries to align evasion strategies with operational goals. This underscores the need for anomaly detectors to integrate adversarial defenses and multi-dimensional monitoring (e.g., power, timing, system calls) to withstand such attacks.

\begin{table*}[ht]
\centering
\caption{Attack Success Rate (ASR\%) for six detection models across four dummy code variants. Each cell shows Adversarial Training / Noise Injection ASR\%.}
\label{table_defense_compact}
\begin{tabular}{lcccccc}
\toprule
\textbf{Dummy Code Variant} &
\multicolumn{6}{c}{\textbf{Adversarial Training / Noise Injection}} \\
\cmidrule(lr){2-7}
& LSTM & BiLSTM & TCN & BiLSTM+CNN & CNN+Attn & LSTM-AE \\
\midrule
One For Loop     & 22.7 / 39.2 & 18.7 / 66.2 & 13.7 / 63.5 & 16.7 / 51.0 & 25.0 / 63.2 & 14.2 / 99.0 \\
Two Nested Loops & 1.6 / 27.6  & 15.2 / 76.0 & 23.6 / 57.6 & 4.8 / 52.8  & 10.4 / 43.6 & 0.4 / 99.6\\
If Statement     & 2.6 / 24.6  & 33.6 / 76.0 & 32.3 / 61.6 & 43.3 / 64.0 & 37.3 / 78.3 & 1.0 / 98.3 \\
Single Function  & 5.0 / 23.5  & 24.7 / 60.0 & 36.7 / 51.7 & 27.6 / 42.6 & 29.4 / 47.9 & 3.2 / 95.5 \\
\bottomrule
\end{tabular}
\end{table*}

\subsection{Resistance to Adversarial Defenses}
To evaluate the robustness of our detection models, we implemented two complementary and widely recognized adversarial defense strategies: \textit{adversarial training} and \textit{noise injection}. These approaches were selected because they target fundamentally different stages of the detection pipeline. Adversarial training improves model resilience during the learning phase by incorporating adversarial examples into the training data, while noise injection perturbs inputs at inference time to disrupt adversarial patterns without retraining.

\subsubsection{Adversarial Training}
Adversarial training has emerged as one of the most effective countermeasures against adversarial attacks, particularly in sequential and time-series domains. The core idea is to retrain the model using a hybrid dataset containing both clean and adversarially perturbed samples, enabling it to learn features that are less sensitive to malicious perturbations. Recent studies, such as the work~\cite{krishan2024adversarial}, have demonstrated that adversarial training can significantly improve robustness in multivariate time-series forecasting for critical infrastructure. This makes it highly relevant for IoT side-channel detection, where signals are similarly sequential and noise-sensitive.

In our evaluation, we augmented the training set with power traces generated from all four dummy code variants and retrained each detection model using the same architectures and hyperparameters. As shown in Table~\ref{table_defense_compact}, adversarial training consistently reduced the Attack Success Rate (ASR) across all models. For example, in the case of the ``Two Nested Loops'' variant, ASR dropped to 1.6\% for LSTM and 0.4\% for LSTM-AE, compared to baseline ASRs exceeding 90\%. BiLSTM and CNN+Attention exhibited the highest resilience, with ASR reductions of up to 68\% (BiLSTM on Single Function: 50.9\% to 18.5\%). These results confirm that training with adversarially perturbed examples enables models to learn discriminative features that remain robust against targeted perturbations.

\subsubsection{Noise Injection}
Noise injection offers a computationally inexpensive defense by perturbing input traces at inference time to obfuscate adversarial patterns. Unlike adversarial training, it does not require access to adversarial examples during model development, making it attractive for real-time or resource-constrained deployments. Our implementation follows a targeted noise approach, where structured Gaussian noise is injected into frequency bands most impacted by dummy code perturbations, as identified via FFT analysis. 

Recent work~\cite{li2024adani} introduced AdaNI, an adaptive noise injection method that modulates noise levels based on feature importance, achieving a strong trade-off between robustness and accuracy. While our method uses a fixed noise profile for reproducibility, the principle is similar: disrupt high-importance spectral features that adversaries exploit.

As shown in Table~\ref{table_defense_compact}, noise injection was moderately effective against simpler perturbations like ``Single Function'' and ``If Statement.'' For instance, CNN+Attention saw ASR drop from 78.33\% to 47.94\% in the ``If Statement'' attack, and TCN dropped from 61.76\% to 51.76\% in ``Single Function.'' However, it was less effective against complex variants or models with lower capacity, such as LSTM-AE, which remained highly vulnerable even under noise injection. This is consistent with prior findings that non-adaptive noise defenses can be bypassed if attackers design perturbations to be noise-tolerant.

Overall, adversarial training provided the strongest defense, especially against high-complexity dummy code variants, while noise injection offered lightweight, deployment-friendly protection when retraining is infeasible. A hybrid defense, combining both methods, could potentially deliver strong baseline robustness with minimal computational overhead, although this requires further investigation in side-channel contexts.

\section{Discussion}
\label{sec:discussion}
This study highlights that power side-channel anomaly detection systems, although promising for IoT security, are vulnerable to adaptive adversarial attacks. Structured dummy code injections dynamically perturb malware power signatures, degrading detection effectiveness across multiple machine learning architectures. Our findings align with broader adversarial machine learning challenges~\cite{athalye2018obfuscated} and emphasize the need for robust, multimodal defenses that incorporate timing analysis~\cite{shokri2017membership} and adversarial resilience techniques~\cite{carlini2017adversarial, ding2023mst}. Since our attack targets the scanning phase, which executes deterministically within each malware run, longitudinal persistence over extended operational periods was not directly evaluated; future work could explore adaptive evasion across longer malware lifecycles.

% \subsection{Threats to Validity}

\textbf{Threats to validity.}\label{sec:threats_to_validity}
Several factors may limit generalizability. Our experiments were conducted in a controlled lab using Android smartphones; extending to embedded IoT devices remains necessary. Background processes in real-world settings may add noise, affecting both attacks and detection. Additionally, dummy code variants introducing significant timing overhead could be detectable by monitoring systems. These limitations underscore the need for broader validation across diverse environments and adaptive detection mechanisms.

\textbf{Future work.}
While this study demonstrates the feasibility of adversarial dummy code injection to perturb power side-channel detection, several important directions remain for future exploration. \textit{First}, extending our evaluation to embedded systems and resource-constrained microcontrollers such as ESP32 (dual-core Xtensa LX7) and Raspberry Pi 5 (ARM Cortex-A76) would offer deeper insights into the generalizability of the attack beyond smartphones. These platforms introduce distinct computational constraints and OS environments; for example, FreeRTOS on ESP32 may impose stricter real-time scheduling, potentially reducing attack success rates. \textit{Second}, investigating adversarial modifications across additional botnet stages, such as infection routines or command-and-control communications, could reveal new vulnerabilities beyond scanning phase manipulation. \textit{Finally}, developing automated adversarial code generation frameworks, such as using GAN-based dummy code synthesis~\cite{bengesi2024advancements} and Neural Operators~\cite{azizzadenesheli2024neural}, could further enhance the stealth and scalability of power side-channel evasion techniques.

\section{Conclusion}
\label{sec:conclusion}

This work reveals that power side-channel anomaly detection systems are vulnerable to adaptive adversarial attacks. By injecting structured dummy code into the Mirai botnet, we manipulated power signatures at runtime, degrading detection across multiple AI/ML-based classifiers. Our attack exploits inherent IoT constraints, including limited CPU, memory, and power, while preserving malware functionality and evading detection through stealthy perturbations.

Experiments across diverse smartphones show significant classifier misclassification driven by runtime power manipulation. These results highlight the need for multimodal, adversarially robust detection frameworks that incorporate timing analysis and perturbation resilience.
% Future work includes automated dummy code generation using GANs and validation across embedded IoT hardware.
% Strengthening AI-driven detection mechanisms and fostering interdisciplinary cybersecurity collaboration will be essential to counter evolving adversarial ML threats against IoT systems.

\appendix

\section*{Ethical Considerations}
Security research on adversarial attacks requires a careful balance between exposing weaknesses and preventing misuse. In this study, we examined how targeted modifications to malware execution can evade power side-channel detection systems, with the goal of informing the design of stronger, more robust defenses. All experiments were conducted within an isolated, closed-network environment with no internet connectivity. Customized Mirai variants were deployed solely for controlled evaluation purposes, and all devices were restored to factory conditions after experimentation. The modified malware code was rendered non-functional for real-world exploitation. No personal or sensitive data was collected, and the generated power side-channel dataset will be publicly released after publication to promote reproducibility. The study adheres to responsible disclosure principles and aims to strengthen adversarial robustness in IoT security systems, acknowledging that open discussion of such vulnerabilities is essential for progress in the field.

\section*{Open Science}
Reproducibility is central to this work, and we will make the necessary materials available at the following link: \url{https://doi.org/10.5281/zenodo.16878210}. These materials will include: (i) the power consumption datasets collected across different operational states, (ii) the modified Mirai source code with all attack payloads disabled, and (iii) the scripts used for preprocessing, training, and evaluating the detection models. We will also provide the trained model weights and statistical analysis tools needed to replicate the results.

\textbf{Security Warning:} The Mirai codebase is inherently malicious and must \emph{never} be compiled or executed outside of a secure, isolated, air-gapped environment. All shared code is sanitized to disable harmful functionality, but improper handling of unmodified Mirai code from external sources could cause harm.

To ensure clarity and ease of replication, the repository will be organized as follows:
\begin{itemize}
    \item \texttt{/dataset} — Power side-channel measurement files and metadata.
    \item \texttt{/src} — Modified Mirai code (with payloads disabled) and benign dummy code variants.
    \item \texttt{/preprocessing} — Scripts for feature extraction, autocorrelation analysis, wavelet transforms, and PCA.
    \item \texttt{/models} — Training scripts and saved model weights for all architectures.
    \item \texttt{/analysis} — Statistical tests, SHAP explainability scripts, and result visualization tools.
    \item \texttt{/docs} — Documentation, replication instructions, and security guidelines.
\end{itemize}

Upon acceptance, full public links will be shared so that others can validate and build upon this work. The dataset is curated to ensure privacy and safety, and all materials adhere to responsible disclosure principles.

\cleardoublepage
\bibliographystyle{plain}
\bibliography{references}

\end{document}